\documentclass[twocolumn,journal]{IEEEtran}
\usepackage[T1]{fontenc}
\usepackage[latin9]{inputenc}
\usepackage{color}
\usepackage{array}
\usepackage{booktabs}
\usepackage{multirow}
\usepackage{amsmath}
\usepackage{amssymb}
\usepackage{graphicx}
\usepackage{mathtools}
\usepackage[numbers,numbers,sort&compress]{natbib}

\usepackage[ruled,vlined,linesnumbered]{algorithm2e}

\usepackage{graphicx}
\makeatletter

\providecommand{\tabularnewline}{\\}

\SetKwInput{KwInput}{Input}
\SetKwInput{KwOutput}{Output}

\usepackage{algpseudocode}
\usepackage{amsmath,amsfonts,amssymb}
\usepackage{array,booktabs,arydshln}
\usepackage{xcolor}
\usepackage{arydshln}
\usepackage{dsfont} 
\usepackage{float} 

\usepackage[caption=false,font=footnotesize]{subfig}

\newcommand{\lqt}[1]{{\color{black}{#1}}}

\newcommand{\dcoflow}{$\mathtt{DCoflow}$}
\newcommand{\wdcoflow}{$\mathtt{WDCoflow}$}
\newcommand{\wdcoflowDP}{$\mathtt{WDCoflow\text{-}DP}$}
\newcommand{\csmha}{$\mathtt{CS\text{-}MHA}$}
\newcommand{\csdp}{$\mathtt{CS\text{-}DP}$}
\newcommand{\sincronia}{$\mathtt{Sincronia}$}
\newcommand{\varys}{$\mathtt{Varys}$}
\newcommand{\cdslp}{$\mathtt{CDS\text{-}LP}$}
\newcommand{\cdslpa}{$\mathtt{CDS\text{-}LPA}$}

\usepackage{tikz}
\usepackage{tcolorbox}

\@ifundefined{showcaptionsetup}{}{%
 \PassOptionsToPackage{caption=false}{subfig}}
\usepackage{subfig}
\makeatother

\newcommand{\cC}{{\cal C}}
\newcommand{\cF}{{\cal F}}

\newcommand{\cL}{{\cal L}}
\newcommand{\cO}{{\cal O}}
\newcommand{\cS}{{\cal S}}
\newcommand{\cR}{{\cal R}}

\newcommand{\hv}{\hat{v}}
\newcommand*\diff{\mathop{}\!\mathrm{d}} 

\usepackage{amsthm}
\newtheorem{theorem}{Theorem}[]
\newtheorem{lemma}[theorem]{Lemma}

\begin{document}
\title{Weighted Scheduling of Time-Sensitive Coflows}


\author{Olivier~Brun, 
    Rachid~El-Azouzi,
    Quang-Trung~Luu,
    Francesco~De~Pellegrini,\\
    Balakrishna~J.~Prabhu,
    and~C\'edric~Richier
\thanks{Parts of this work have been presented at IFIP Networking 2022.}
\thanks{Olivier~Brun and Balakrishna~J.~Prabhu are with LAAS-CNRS, University of Toulouse, CNRS, 31400  Toulouse, France (e-mails:
\{brun, bala\}@laas.fr).}
\thanks{ Rachid~El-Azouzi, Francesco~De~Pellegrini, and C\'edric~Richier are with CERI/LIA, University of Avignon, 84029 Avignon, France (e-mails: \{rachid.elazouzi, francesco.de-pellegrini, cedric.richier\}@univ-avignon.fr).}
\thanks{Quang-Trung Luu is with the School of Electrical and Electronic Engineering, Hanoi University of Science and Technology, 100000 Hanoi, Vietnam (e-mail: trung.luuquang@hust.edu.vn).}
}

\markboth{Preprint}%
{Brun \MakeLowercase{\textit{et al.}}: Weighted Scheduling of Time-Sensitive Coflows}



\maketitle


\begin{abstract}
Datacenter networks commonly facilitate the transmission of data in distributed computing frameworks through coflows, which are collections of parallel flows associated with a common task. Most of the existing research has concentrated on scheduling coflows to minimize the time required for their completion, i.e., to optimize the average dispatch rate of coflows in the network fabric. Nevertheless, modern applications often produce coflows that are specifically intended for online services and mission-crucial computational tasks, necessitating adherence to specific deadlines for their completion. In this paper, we introduce  \wdcoflow,~ a new algorithm to maximize the weighted number of coflows that complete before their deadline. By combining a dynamic programming algorithm along with parallel inequalities, our heuristic solution performs at once coflow admission control and coflow prioritization, imposing a $\sigma$-order on the set of coflows. With extensive simulation, we demonstrate the effectiveness of our algorithm in improving up to $3\times$  more  coflows that meet their deadline in comparison the best SoA solution, namely $\mathtt{CS\text{-}MHA}$. Furthermore, when weights are used to differentiate coflow classes, \wdcoflow~ is able to improve the admission per class up to $4\times$, while increasing the average weighted coflow admission rate.
\end{abstract}

\IEEEpeerreviewmaketitle{}


\section{Introduction
\label{sec:Introduction}}


\IEEEPARstart{T}{he} concept of coflow, firstly introduced in \cite{ChowdhuryHotNet2012}, forms the foundation of modern traffic engineering in datacenter networks. This abstraction of traffic was initially developed to capture the patterns of data exchange within distributed computing frameworks like MapReduce or Spark \cite{dean2004mapreduce, zaharia2010spark}. These frameworks employ the \textit{dataflow} computing model for processing large-scale data, which involves distributing intermediate computation stages across multiple nodes and transferring outputs to nodes responsible for the subsequent stages. During the transitions between computation stages, dataflows generate a set of network flows that traverse the datacenter fabric. These flows are abstracted as a \textit{coflow}. A prominent example of a dataflow occurs in the \textit{shuffle phase} of the Hadoop MapReduce framework \cite{dean2004mapreduce}.
 However, it has been investigated  in real traces \cite{chowdhury2011managing} that coflow scheduling has a significant impact on the completion time of applications and the  \textit{shuffle phase}  accounts for 33\% of the running time in observed coflows.   Hence, the reference objective function to measure acceleration at network layer  is the makespan or Weighted Coflow Completion Time (WCCT). Minimizing the average WCCT or CCT is an appropriate objective for maximizing the number of computing jobs dispatched per hour in a datacenter fabric. 
Numerous works, such as \cite{ChowdhuryHotNet2012,ChowdhuryThesis2015,Shaf2018,agarwal2018sincronia,ELITE22, ahmadi2020scheduling, Shafiee2018improved}, have addressed the minimization of  WCCT and proposed algorithmic solutions. Over the past decade, extensive research has illuminated the complexity of this problem. It has been proven to be {\em NP}-hard and inapproximable below a factor of  $2$ through reduction to the job scheduling problem on multiple correlated machines. Near-optimal methods have also been proposed in the literature, with performance bounds approximating a factor of  $4$ \cite{Shaf2018,agarwal2018sincronia,chowdhury2019near}.
However, the context radically changes when dealing with time-critical jobs that impose strict deadlines on the coflow's data transfer phase.


In such scenario, the scheduling of coflows is commonly combined with admission control to minimize the number of deadline violations, i.e., the number of coflows that are unable to be completely transferred before their deadlines. This gives rise to the Coflow Deadline Satisfaction (CDS) problem, first introduced in \cite{Tseng2019}. Each coflow is assigned a specific deadline, and the objective is to perform joint \textit{coflow admission control and scheduling} to maximize the number of admitted coflows that can meet their respective deadlines. This problem is also proven to be {\em NP}-hard, and it has been shown to be inapproximable within any constant factor of the optimal solution \cite{Tseng2019}.


While the issue of time-sensitive coflows has been acknowledged in the early literature \cite{Chowdhury2014}, most works on coflow scheduling have not focused on addressing this problem, with a few exceptions \cite{Tseng2019}. However, our performance analysis has revealed that even near-optimal algorithms designed for minimizing CCT may fail to meet coflow deadlines. In reality, the concept of time-sensitive coflows has become increasingly prevalent in modern distributed datacenters. It is not only computing frameworks that deal with time-sensitive tasks; modern web and mobile applications are built using microservice architectures, where user requests can trigger numerous services across multiple servers to retrieve data. The completion time of a batch of flows, i.e., the time instant at which the last bit of data arrives, determines the lag to the response time of these services, and significant delays can lead to a degraded user experience. In the realm of cloud computing and data centers, there is a rise in more time-sensitive applications, such as web search \cite{web} and machine learning \cite{Jiacheng19}, which impose stricter deadline constraints. With this performance objective in mind, a coflow is only considered beneficial when all of its individual flows have completed their data transfer within the required deadline.

 To reduce the number of coflows missing their deadlines, i.e., the number of violations, existing solutions perform admission control. 
 On the other hand, in solving the CDS problem one has to operate simultaneously both {\em coflow admission control and scheduling}. This allows to maximize the number of admitted flows while respecting their deadlines, i.e., the Coflow Acceptance Rate (CAR).

 In this paper, we generalize the CDS problem to the case when coflows have a priority in the form of a nonnegative weight. The performance metric to maximize is the Weighted Coflow Acceptance Rate (WCAR). Since maximizing CAR is {\em NP}-hard \cite{Tseng2019}, the same is true for maximizing WCAR and exact solution methods are of little practical use. 
In principle, it is possible to address the problem of time-sensitive coflows by formulating a suitable Mixed Integer Linear Program (MILP). However, when dealing with datacenters that handle tens of thousands of coflows \cite{ChowdhuryThesis2015}, techniques relying on MILPs or their relaxations may not be practical or feasible. The computational complexity and scalability challenges associated with solving MILPs in such large-scale environments make them less viable for real-time implementation.
 

To achieve scalability in coflow scheduling for datacenters, the use of scalable algorithms is crucial. Many research works propose the concept of scheduling coflows using a priority order, known as the $\sigma$-order. Once the $\sigma$-order is determined, a work-conserving transmission policy can be adopted. The focus on $\sigma$-order schedulers is driven by their implementation advantages. Specifically, in terms of rate control, any work-conserving preemptive dynamic rate allocation can be used as long as it is compatible with the assigned coflow priorities. It has been shown that the maximum performance loss within such rate allocation policies is bounded by a factor of $2$ \cite{agarwal2018sincronia}. For example, using fixed coflow priorities under DiffServ satisfies the definition of a $\sigma$-order scheduler. Additionally, commercial switches often have built-in priority queues and support per-flow tagging, which can be utilized to prioritize active coflows without requiring per-flow rate control. This allows for a greedy rate allocation that aligns with the desired $\sigma$-order. The exact mapping between a coflow's $\sigma$-order and the switch's priority queuing mechanism, as well as the limitations imposed by legacy hardware, are interesting subjects but beyond the scope of this paper.

\noindent{\it Contributions.} In this paper, we introduce lightweight algorithms for coflow scheduling with deadlines. The proposed algorithms surpass existing solutions in the literature and do not rely on solving linear programs. The proposed approach consists of an offline admission control policy combined with a scheduler belonging to the class of $\sigma$-order coflow schedulers. The output of the algorithm is a priority order restricted to the set of admitted coflows. Our heuristic solutions, named \wdcoflow, leverage techniques such as  dynamic programming  \cite{Lenstra1976} (known for optimality in the single link case) and parallel inequalities for completion times \cite{Schulz1996}. By employing these techniques, our heuristics effectively capture the inter-coflow impacts and determine the coflows that should be admitted. The algorithms are further extended to handle joint admission control and scheduling in online scenarios where coflows are generated at runtime with unknown release times.
Through extensive numerical experiments on various scenarios, including both synthetic and real traces obtained from the Facebook data \cite{Chowdhury2014}, we demonstrate that our algorithm consistently outperforms existing solutions in the literature. The simulations encompass offline and online settings, and the \wdcoflow\ algorithm consistently achieves near-optimal Weighted Coflow Arrival Rate (WCAR) for smaller fabrics. Moreover, it outperforms state-of-the-art solutions across all evaluated workloads, achieving significant improvements of up to 4 times in certain cases, especially for overloaded fabrics.





The remaining sections of the paper are structured as follows.  Sec.~\ref{sec:Problem-Statement} presents an overview of the general problem addressed in the paper, including the description of coflow ordering models.  Sec.~\ref{sec:heuristic} introduces the proposed algorithms, detailing their design and methodology. Numerical results are then presented in Sec.~\ref{sec:Evaluation}.  Sec.~\ref{sec:Related-Work} discusses the related work in the field of scheduling time-sensitive coflows. Finally,  Sec.~\ref{sec:Conclusion} presents concluding remarks and outlines potential directions for future research.



\section{Problem Statement}
\label{sec:Problem-Statement}



In this section, we  formally define the deadline scheduling problems for coflows with weights as an MILP.
\begin{figure}[t]
\centering
\includegraphics[width=0.65\columnwidth]{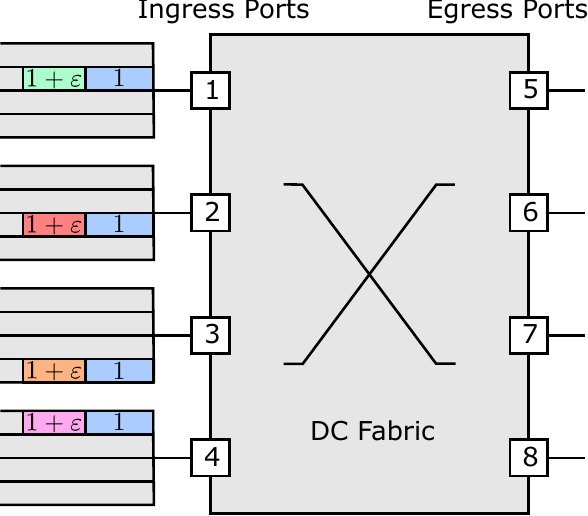}
\caption{Example of a Big-Switch fabric having $4$ ingress/egress ports connecting to $4$ machines. Flows in ingress ports are organized by destinations and are color-coded by coflows. The example has $5$ coflows. \lqt{Coflow $k_1$ (blue) has $4$ flows, with each ingress port sending $1$ units of data to one egress port: its deadline is $1$;
coflows $k_2$ (green), $k_3$ (red), $k_4$ (orange) and $k_5$ (purple) have a single flow, each sending $(1+\varepsilon)$ unit of data}. The deadline of these coflows is $2$.  \label{fig:Example}}
\end{figure}

The datacenter network is modeled as a non-blocking switch, as in Fig. \ref{fig:Example}. This is usually referred to as the  \textit{Big-Switch} model, used for the first time for datacenter coflow scheduling in \cite{Chowdhury2014}. In that model, two disjoint sets of ports, namely the {\em ingress} ports and the {\em egress} ports, represent all the ports of the Top of Rack (ToR) switches connecting machines hosted in racks to the network fabric. The mathematical model for a switch is set of ports (or links) $\cL=\{1,2,\hdots, 2M\}$ where $\ell \in \{1,\hdots,M\}$ are ingress ports and $\ell \in \{M+1,\hdots,2M\}$ are egress ports. We assume that each port $\ell\in\cL$ has a maximum rate of $B_\ell$. 

 A coflow is a set of flows, where each flow is a volume of data to be transferred between an ingress port and an egress port. In the example of Fig. \ref{fig:Example}, at each ingress port, flows are organized in virtual output queues indicating the output port.

For clarity, the scheduling problem is formulated in the offline setting. Hence, all the coflows are available at time $0$, i.e., when the scheduling decision is taken. Later on, the algorithms will be adapted for the online setting where scheduling decision are taken over the course of a given time horizon, and coflows arrive over time. In turn, the characteristics of the future coflows are unknown.

Consider a batch of $N$ coflows $\cC = \{1, 2, ..., N\}$.  We denote by $w_k$ the weight, i.e., the importance, of coflow $k$, so that the acceptance rate can be optimized with regard to its weight.  Each coflow $k$ is subject to a completion deadline $T_k$. 
The set of flows of coflow $k$ is denoted by $\cF_{k}$. A flow is defined by its volume and the pair of ports that it uses. Let $v_{k,j}$ be the volume of flow $j$ of coflow $k$, and let $\cF_{\ell,k}$ be the set of flows in $\cF_{k}$ that uses port $\ell \in \cL$ either as ingress port or as egress port.   Table~\ref{tab:Notations} summarizes the main notations used throughout the paper.

\begin{table}[t]
\caption{Main notations. \label{tab:Notations}}
\footnotesize
\centering
\begin{tabular}{cl}
\toprule 
\textit{Symbol} & \textit{Description}\tabularnewline
\cmidrule[0.4pt](lr{0.12em}){1-1}%
\cmidrule[0.4pt](lr{0.12em}){2-2}%
 $\mathcal{L}$ & set of fabric ports\tabularnewline
$B_{\ell}$ & available bandwidth of port $\ell\in\mathcal{L}$\tabularnewline
$\mathcal{C}$ & set of coflows, $\left|\mathcal{C}\right|=N$\tabularnewline
$v_{k}$ & volume of coflow $k$\tabularnewline
$w_{k}$ & weight of coflow $k$\tabularnewline
$T_{k}$ & deadline of coflow $k$\tabularnewline
$z_{k}$ & binary indicator for the admission of coflow $k$\tabularnewline
$c_{k}$ & completion time of coflow $k$\tabularnewline
$c_{\ell,k}$ & completion time of coflow $k$ on port $\ell$\tabularnewline
$\mathcal{F}_{k}$ & set of flows of coflow $k$\tabularnewline
$\mathcal{F}_{\ell,k}$ & set of flows of coflow $k$ that use port $\ell$\tabularnewline
$v_{k,j}$ & volume of flow $j\in\mathcal{F}_{k}$\tabularnewline
$p_{\ell,k}$ & processing time of coflow $k$ on port $\ell$\tabularnewline
$\sigma$ & scheduling order of coflows, $\sigma=\left\{ \sigma_{1},\cdots,\sigma_{N-1},\sigma_{N}\right\} $\tabularnewline
$\mathds{1}_{k,i}$ & binary indicating whether coflow $k$ is of class or not\tabularnewline
$P^{\left(j\right)}\left(w\right)$ & minimum total processing time for any feasible subset \tabularnewline
 & of coflows $\left\{ 1,\cdots,j\right\} $ that has total weight $w$\tabularnewline
\bottomrule
\end{tabular}
\end{table}

\subsection{MILP Formulation}
\label{ssec:milp}

Let $z_{k} \in \{0,1\}$ be an indicator of whether coflow $k$ finishes before its deadline $T_k$, and let $r_{k,j}(t) \in \mathbb {R}_{+}$ be the rate allocated to flow $j\in {\cal F}_k$ at time $t$. 
The target coflow scheduling problem prescribes to identify the set of coflows to be scheduled in order to maximize the corresponding cumulative weight. 
We will refer to this scheduling problem as \textit{Weighted Coflow Acceptance Rate} (WCAR) problem, which is formulated as
\begin{align}
    \underset{r}{\mathrm{max}}\enskip & \sum_{k\in\mathcal{C}}  w_kz_{k} \tag{WCAR}\label{prob:WCAR}\\
    \mathrm{s.t.}\enskip 
    & \sum_{k \in \mathcal{C}} \sum_{j \in \mathcal{F}_{\ell,k}} r_{k,j}(t)\leq  B_{\ell}, \quad \forall \ell \in \mathcal{L}, \forall t \in \mathcal{T}, \label{prob:WCAR-1} \\
    & \int_0^{T_k} r_{k,j} (t) \diff{t} \geq  v_{k,j} z_k, \quad \forall j \in \mathcal{F}_k, \forall k \in \mathcal{C}, \label{prob:WCAR-2}
\end{align}
%
where $\mathcal{T}$ is the time interval over which scheduling is performed and can be set to $[0, \max_k T_k]$. Constraint \eqref{prob:WCAR-1} ensures that the total rate allocated on port $\ell$ at every time instant in $\mathcal{T}$ does not exceed its capacity $B_{\ell}$. Constraint \eqref{prob:WCAR-2} ensures that all flows of every \textit{accepted} coflow are completely processed before the deadline of that coflow. Note that solving the \ref{prob:WCAR} problem requires to define optimal rate allocations $r_{k,j} (t)$,  $\forall j \in \mathcal{F}_k$, $\forall k \in \mathcal{C}$, and $\forall t \in \mathcal{T}$.

Assume without loss of generality that coflows are numbered in the Earliest Due Date (EDD) order. It is then clear that, given a feasible schedule, only coflows $\{k+1, \ldots, N\}$ are present in the system in time interval $[T_k, T_{k+1}]$. Assuming that the rate allocations $r_{k,j} (t)$ are constant in the time intervals $[0, T_1], [T_1, T_2], \ldots, [T_{N-1},T_N]$, we obtain a MILP formulation of Problem \eqref{prob:WCAR}, which generalizes the formulation proposed in \cite{Tseng2019} for unweighted coflows.

\lqt{
When all the coflows have the same weight, it was shown in \cite{Tseng2019} that Problem~\eqref{prob:WCAR} is \textit{NP}-hard.
\begin{lemma}[Proposition 1 in \cite{Tseng2019}]
\label{lemma1}
When the weights are equal, there exists a polynomial time reduction of Problem~\eqref{prob:WCAR} to the problem of minimizing the number of late jobs in a concurrent open shop \cite{Lin2007}. Hence, Problem~\eqref{prob:WCAR} is \textit{NP}-hard. 
\end{lemma}
For completeness, we restate the result for unequal weights as well although it is direct consequence of the problem with equal weights.
}

\subsection{Upper Bound ILP for WCAR}\label{ssec:lb}


Problem~\eqref{prob:WCAR} solves for the rate allocation and determines which coflows satisfy their deadline. It thus allows rate allocations that share ports' capacity possibly among several coflows. An alternative approach is to determine an ordering $\sigma$ of coflows first and then assign full port rates to coflows that have higher priority according to $\sigma$. Hence, flow $j \in \cF_{\sigma_{k}}$ is blocked if and only if either its ingress or egress port is busy serving a flow $j' \in  \cF_{\sigma_{k'}}$ for some $k' < k$ in the $\sigma$-order. The order thus implies a strict priority on the ports utilization. A flow scheduling that follows this priority rule is called \emph{$\sigma$-order-preserving}. 

The coflow ordering approach was first taken in \cite{agarwal2018sincronia} for the minimization of Coflow Completion Times (CCT). It was then applied to deadline scheduling but without weights in \cite{Dcoflow2022}. The advantage of this approach is that it does not require rate computations. Once an order is determined, the rates can be deduced directly from there. On the other hand, it has the disadvantage of being an upper bound for deadline scheduling as shown in \cite{Dcoflow2022}.

Here, we give a short summary of those arguments. The problem of finding the optimal $\sigma$-order is in fact an ILP. To see this, we will need to define a couple of terms. The processing time in isolation of coflow $k$ at port $\ell$ is defined as 
$p_{\ell,k} = \hv_{\ell,k} / B_\ell$,
where $\hv_{\ell,k}=\sum_{j \in \cF_{\ell,k}} v_{k,j}$ is the total volume sent by coflow $k$ on port $\ell$. That is, $p_{\ell,k}$ is the time to transfer all the data of coflow $k$ on port $\ell$ in the absence of other coflows. Further, for $k'\neq k$, define the binary variable $\delta_{k',k}$ which is $1$ if coflow $k'$ has a higher priority that $k$. and $0$ otherwise. An ordering $\sigma$ can then be derived from the variables $\{ \delta_{k,k'} \}_{k,k' \in \cC}$ by subjecting them to the standard disjunctive constraints
\begin{eqnarray}
    \delta_{k,k'} + \delta_{k',k} & = & 1, \quad \forall k,k' \in \cC, \label{eq:order-disjunct-1} \\
    \delta_{k,k'} + \delta_{k',k"} + \delta_{k",k} & \leq & 2, \quad \forall k,k',k" \in \cC. \label{eq:order-disjunct-2}
\end{eqnarray}

The only step remaining now is to express the constraint that accepted coflows should have a CCT smaller than their deadline in a linear form. \lqt{Unfortunately, there are no known linear inequalities to express the region of schedulability of coflows in a switch}. The difficult arises from the blocking nature of the switch: a flow may be blocked because either its ingress or egress port is being used by another flows. Therefore, transmission times on a port depend on what happens on the other ports.

Nevertheless, the following lower bound on the completion time of coflow $k$ on port $\ell$, $c_{\ell,k}$ can be obtained by assuming the ports are independent,
\begin{equation} \label{eq:lower-bound-CCT}
    c_{\ell,k} \geq \sum_{k' \neq k} p_{\ell,k'} \delta_{k',k} z_{k'} + p_{\ell,k}z_k, \quad \forall \ell \in \cL, k \in \cC.
\end{equation}
Here, only accepted coflows, i.e. those for which $z_k=1$, are accounted for in the bound \eqref{eq:lower-bound-CCT} (hence the term $p_{\ell,k}z_k$). The lower bound on $c_{\ell,k}$ is then just the time it takes to transmit all the coflows with priority higher than $k$ on port $\ell$. 
The product $\delta_{k',k} z_{k'}$ can easily be linearized by introducing binary variables $y_{k',k}$ satisfying the constraints 
\begin{equation}
    y_{k',k} \leq z_{k'}; 
    \quad y_{k',k} \leq \delta_{k',k}; 
    \quad y_{k',k} \geq z_{k'}+\delta_{k',k} - 1. \label{eq:y1} 
    \end{equation}
The lower bound \eqref{eq:lower-bound-CCT} can now be rewritten as the following linear inequality:
    \begin{equation} \label{eq:lower-bound-CCT-new}
    c_{\ell,k} \geq \sum_{k' \neq k} p_{\ell,k'} y_{k',k} + p_{\ell,k}z_k, \quad \forall \ell \in \cL, k \in \cC.  
\end{equation}
Since the CCT of coflow $k$ is given by $c_k =\max_{\ell \in \cL} c_{\ell,k}$, the constraint that the CCT of this coflow is smaller that its deadline can be expressed as
\begin{equation} \label{eq:deadline-constraint}
    c_{\ell,k} \leq T_k z_k, \quad \forall \ell \in \cL, k \in \cC.  
\end{equation}

\lqt{
Finally, the optimal $\sigma$-order coflow scheduling problem can be formulated as the following  ILP,
\begin{equation} \label{prob:COWCAR}
    \mathrm{max}\enskip \sum_{k\in\mathcal{C}} w_kz_{k}\tag{$\sigma$-WCAR}, \quad
    \mathrm{s.t.}\enskip (\ref{eq:order-disjunct-1}, \ref{eq:order-disjunct-2}, \ref{eq:y1}, \ref{eq:lower-bound-CCT-new}, \ref{eq:deadline-constraint}).\nonumber
\end{equation}
Recall that solutions of Problem~\eqref{prob:COWCAR} provide an upper bound on the number of accepted coflows to that of Problem~\eqref{prob:WCAR}.
}
%

\subsection{Motivating Example}
\label{subsec:Motivating-Example}

\lqt{
We now illustrate, with an example, some of the shortcomings of \csmha~\cite{Luo2016}, an algorithm for maximizing the acceptance ratio of coflows \textit{without} weights (i.e., maximizing the CAR).
$\mathtt{CS\text{-}MHA}$  introduces a novel approach to solve the scheduling problem by employing a static coflow prioritization. This prioritization is utilized to approximate the solution for the coflow scheduling problem that maximizes the CAR. 
First, $\mathtt{CS\text{-}MHA}$ computes the scheduling order  and the set of admitted coflows at each port using the Moore-Hodgson algorithm \cite{Moore1968}. Since different ports may have different sets of admitted coflows, a coflow is admitted only if it is admitted by all ports simultaneously.
Then, for the coflows that are rejected, a second round is conducted to reassess if some of them can actually meet their deadlines. In this case, $\mathtt{CS\text{-}MHA}$ selects the coflow with the minimum bandwidth requirement at the bottleneck port. This choice is based on the reasoning that coflows with lower bandwidth requirements are more likely to catch up with their deadlines.
}

\lqt{
Fig.~\ref{fig:Example} shows a simple example to illustrate the shortcomings of  \csmha. This will be used as a running example throughout the paper. The example consists of five coflows: $k_1$ with four flows, and $k_2, k_3, k_4,$ and $k_5$ with one flow each. To facilitate the presentation, the flows are organized in virtual output queues at the ingress ports, where the virtual queue index represents the flow output port modulo the number of machines. The numbers on the flows' representations indicate their \textit{normalized} volumes. All fabric ports have the same \textit{normalized} bandwidth of $1$.

In the first iteration,  \csmha\  uses the Moore-Hodgson algorithm to compute the scheduling order at each port, as mentioned earlier. This algorithm is based on the EDD rule with objective to minimize the number of missed deadlines on a single machine (or port in the coflow context).
In this example, since coflow $k_1$ uses all ports and has the smallest deadline ($T_1 = 1$), it will be scheduled first at each port. Consequently, all other coflows are rejected because they cannot meet their deadlines when scheduled after $k_1$. This results in a coflow scheduling with a CAR of $\frac{1}{5}$.
However, an optimal scheduling solution would be ${k_2, k_3, k_4, k_5, k_1}$ or any combination where coflow $k_1$ is scheduled last. This scheduling achieves a CAR of $\frac{4}{5}$.

To further illustrate the limitations of  of the \csmha, consider now the case where there are $M$ machines, coflow $k_1$ utilizes all ports, and coflows $k_2, \ldots, k_M$ have one flow each. The other parameters remain unchanged.
In this setting, we shall demonstrate that the CAR obtained using \csmha\ and \dcoflow\footnote{\dcoflow\ \cite{Dcoflow2022} is the variant of \wdcoflow\ that deals with unweighted coflows. Detailed differences between \dcoflow\ and \wdcoflow\ shall be given in Sec.~\ref{subsec:dcoflow-offline}.} are respectively $\frac{1}{M}$ and $\frac{M-1}{M}$. With this setting, when the $M$ increases, \csmha\  yields a CAR close to zero, while with \dcoflow, it is close to one.

The key observation in this example is that how \csmha\  neglects the impact that a coflow may have on other coflows across multiple ports. Specifically, a coflow that leads to the missing of multiple deadlines should have a lower priority, even if its own deadline is the earliest. Neglecting this consideration leads to a misjudgment in the coflow ordering, resulting in a final schedule that significantly degrades the CAR compared to an optimal solution.
Building upon this observation, in what follows, we propose  a new class of $\sigma$-order schedulers called \wdcoflow\ to address the joint coflow admission control and scheduling problem.
}

\section{$\sigma$-Order Scheduling with \wdcoflow 
\label{sec:heuristic}}

In this section, we present \wdcoflow, an algorithm to solve the problem of joint coflow admission control and scheduling. Given a list of $N$ coflows and their respective weights, it provides a permutation $\sigma=(\sigma_1,\sigma_2,..,\sigma_N)$ of these coflows, with the aim of maximizing the coflow acceptance rate. A key ingredient of our algorithm is a simple rule for deciding which coflow to reject when there is no feasible schedule. This rule is based on a necessary schedulability condition which is established in Sec.~\ref{subsec:coflow-rejection-rule}. We describe our algorithm for solving offline instances in Sec.~\ref{subsec:dcoflow-offline} and Sec.~\ref{subsec:filter}. Finally, the online implementation of \wdcoflow \ is described in Sec.~\ref{subsec:dcoflow-online}.


\subsection{A Necessary Schedulability Condition \label{subsec:coflow-rejection-rule}}
Given a subset $\cS \subseteq \cC$ of \emph{admitted} coflows, a feasible schedule of ${\mathcal S}$ is a processing order of coflows such that $c_k \leq T_k$, $\forall k \in {\mathcal S}$, where $c_k$ represents the completion time of coflow $k$. We establish below a necessary condition for such a schedule to exist and show how it can be used to decide which coflows should be admitted.

Given $\cS \subseteq \cC$ and a coflow $k \in \cS$, let $\cS_k^-$ be the set of coflows in $\cS$ which are scheduled before $k$ (i.e., coflows of higher priority). By assuming that the transmission of coflow $k$ on port $\ell$ can start as soon as all flows of all coflows $j \in \cS_k^-$ have been transmitted on port $\ell$, we can obtain a lower bound on the completion time of coflow $k$

\begin{equation}
c_k \geq p_{\ell,k} + \sum_{j \in \cS_k^-} p_{\ell,j}, \quad \forall  \ell \in \cL.
\label{eq:parallel-inequalities-1}
\end{equation}

Multiplying \eqref{eq:parallel-inequalities-1} on both sides by $p_{\ell,k}$ and summing over all coflows $k \in \cS$ yield


\begin{align} \label{eq:parallel-inequalities}
 & \sum_{k\in\mathcal{S}}p_{\ell,k}c_{k}\geq\sum_{k\in\mathcal{S}}\left(p_{\ell,k}\right)^{2}+\sum_{k\in\mathcal{S}}p_{\ell,k}\sum_{j\in\mathcal{S}_{k}^{-}}p_{\ell,j}\nonumber \\
 & \quad=\frac{1}{2}\sum_{k\in\mathcal{S}}\left(p_{\ell,k}\right)^{2}+\frac{1}{2}\left[\sum_{k\in\mathcal{S}}\left(p_{\ell,k}\right)^{2}+2\sum_{k\in\mathcal{S}}p_{\ell,k}\sum_{j\in\mathcal{S}_{k}^{-}}p_{\ell,j}\!\right]\nonumber \\
 & \quad=f_{\ell}(\mathcal{S}),
\end{align}

\noindent where
\begin{equation}
f_\ell(\cS)= \frac{1}{2}\sum_{k\in\mathcal{S}}\left(p_{\ell,k} \right)^{2}+\frac{1}{2}\left(\sum_{k\in\mathcal{S}}p_{\ell,k}\right)^{2}.
\label{eq:def-f-ell-S}
\end{equation}

\lqt{
From \eqref{eq:parallel-inequalities}, we can conclude that the CCTs $\left \{ c_k \right \}_{k \in \cS}$ necessarily satisfiy the condition $\sum_{k \in \cS} p_{\ell,k} c_k \geq f_\ell(\cS) $ for any port $\ell \in \cL$ and for any subset $\cS \subseteq \cC$ of \emph{admitted} coflows. These conditions are referred to as the \textit{parallel inequalities}, and they serve as valid inequalities for the concurrent open shop problem \cite{Mastrolilli2010}. It is important to note that these inequalities are independent of the coflow ordering and solely depend on the set of admitted coflows.

We now use the parallel inequalities to determine the coflows that should be rejected, if any.  More precisely, given a set $\cS$ of coflows, we define for each port $\ell \in \cL$ the quantity
}
\begin{equation}
I_\ell \left ( \cS \right ) \triangleq \sum_{k \in \cS} p_{\ell,k} T_k - f_\ell(\cS) \geq 0,
\label{eq:schedulability-index}
\end{equation}

\noindent and use it as a measure of the schedulability of the set $\cS$ of flows. Indeed, if $I_\ell \left ( \cS \right )<0$, it follows from \eqref{eq:parallel-inequalities} and \eqref{eq:schedulability-index}  that $\sum_{k \in \cS} p_{\ell,k} T_k < f_\ell(\cS) \leq \sum_{k\in\mathcal{S}}p_{\ell,k} c_k$, which implies that at least one coflow in $\cS$ is late, whatever the order in which these coflows are scheduled. In other words, $I_\ell \left ( \cS \right ) \geq 0$ for all $\ell \in \cL$ is a necessary condition for a feasible schedule of $\cS$ to exist.

The set $\cL^\star=\left \{ \ell \in \cL \ : I_\ell(\cS) < 0 \right \}$ then represents the set of ports on which at least one coflow is late, whatever the order in which the coflows are (locally) processed. Hence, if $\cL^\star \neq \varnothing$, at least one coflow $k^\star$ using one or more ports in $\cL^\star$ should be removed from $\cS$ so as the remaining coflows can meet their deadlines. 

If there is only one port $\ell$ in $\cL^\star$, a natural choice is to choose $k^\star$ so as to maximize the quantity $I_\ell \left ( \cS \setminus \{ k^\star \} \right )$ in the hope that it becomes positive. Observe that for any $j \in \cS$
 
 \begin{align}
f_{\ell}\left(\cS \right) & =  \frac{1}{2} \left [ p_{\ell,j}^{2}+\sum_{k \neq j} p_{\ell,k}^{2}  + \left ( p_{\ell j}+\sum_{k \neq j}p_{\ell k} \right )^{2} \right ]   \nonumber \\
 & =  f_{\ell} \left ( \cS \setminus \{ j \} \right ) + p_{\ell, j} \sum_{k \in \cS} p_{\ell,k}, \label{eq:fls_split}
\end{align}

\noindent from which it follows that $I_\ell \left ( \cS \setminus \{ j \} \right )=I_\ell \left ( \cS \right ) + \Psi_{\ell,j}$, where

\begin{equation}
\Psi_{\ell,j}=p_{\ell,j} \left ( \sum_{k \in \cS} p_{\ell,k} - T_{j} \right ).
\end{equation}

Hence, maximizing $I_\ell \left ( \cS \setminus \{ j \} \right )$ is equivalent to maximizing $\Psi_{\ell,j}$. As coflows with small weights should be rejected in priority, we choose $k^\star \in \mbox{argmax}_{j} \frac{1}{w_j} \Psi_{\ell,j}$. In words, this rule dictates to reject a coflow $k^\star$ with a small weight $w_{k^\star}$ and which has either a large processing time $p_{\ell,k^\star}$ or a large deadline violation $\sum_{k \in \cS} p_{\ell,k} - T_{k^\star}$ when scheduled as the last one, or both.

When there is more than one port in $\cL^\star$, a straightforward extension of  the previous rule is to choose a coflow $k^\star$ with a small weight so as to maximize $\sum_{\ell \in \cL^\star} I_\ell \left ( \cS \setminus \{ k^\star \} \right )$. In this case, we choose the coflow $k^\star$ with the largest value of the index  $\frac{1}{w_k^\star} \sum_{\ell \in \cL^\star} \Psi_{\ell,k^\star}$. An obvious advantage of this simple rule is that it allows to account for the impact of the removal of coflow $k^\star$ on all ports $\ell \in \cL^\star$ used by this coflow.

\subsection{Offline Algorithm \label{subsec:dcoflow-offline}}

The proposed offline algorithm, namely \wdcoflow, is inspired from the \dcoflow\ algorithm proposed in \cite{Dcoflow2022}, which was devised for the unweighted setting. It takes as input a set $\cC=\{1,2,\ldots,N\}$ of coflows, which are all available at time $0$, and computes as output the scheduling order of accepted coflows. The pseudocode of \wdcoflow\ is described in Algorithm~\ref{algo:WDCoflowDP}. We have omitted the $\mathtt{RemoveLateCoflows}$ subroutine in the pseudocode since it is the same as in \cite{Dcoflow2022}. In what follows, we highlight the main steps of \wdcoflow\ and its main differences to  \dcoflow.

\begin{algorithm}[h]
\footnotesize
\caption{\wdcoflow \label{algo:WDCoflowDP}}

\BlankLine

Set $\cS=\left \{ 1,2,\ldots, N \right \}$ and $n = N$; \Comment{initialization}

\While{ $\cS \neq \varnothing$ }{

    Compute $t_{\ell}=\sum_{k \in \cS} p_{\ell, k}$  $\forall\,\ell \in \cL$ and $\ell_b = \underset{\ell \in {\cL}}{\mathrm{arg\,max}}\,  t_{\ell}$; 
    
    Set $\cS_b = \left \{ k \in \cS \ : \ p_{\ell_b, k} > 0 \right \}$; \Comment{coflows in $\cS$ using $\ell_b$}
    
    Set $k' = \underset{k \in {\cS_b}}{\mathrm{arg\,max}}\, T_k$ \Comment{max-deadline coflow on $\ell_b$}
	
	\eIf{$t_{\ell_b} \leq T_{k'}$}
	{
        Set $\sigma_n = k'$ and $\sigma_n^\star=0$ \Comment{admit coflow $k'$}
 
	}{
		
		Set $k^{\star} = \mathtt{RejectCoflow}(\cS_b)$; \Comment{select a coflow to reject}
	
		Set  $\sigma_{n} = k^{\star}$ and $\sigma_{n}^\star = k^{\star}$; \Comment{pre-reject coflow $k^{\star}$}; 
		
	} 
	
	$\cS = \cS \backslash \left\{ \sigma_n \right\}$; \Comment{remove coflow $\sigma_n$ from $\cS$} 

	$n = n-1$; \Comment{update the iteration index} 

} 
$\sigma = \mathtt{RemoveLateCoflows} \left( \sigma, \sigma^{\star} \right)$; 

\KwRet $\sigma$;  \Comment{final scheduling order}

\vspace{0.1cm}
\SetKwFunction{FMain}{RejectCoflow} 
\SetKwProg{Fn}{Function}{:}{} 


\Fn{\FMain{\textit{$\mathcal{S}_b$}}}{

Set $\mathcal{R} = \mathtt{Filter}(\cS_b)$ \Comment{candidate coflows for rejection}

Set $k^{\star} = \arg\max_{j\in\mathcal{R}}\frac{1}{w_j}  \sum_{\ell \in \cL^\star} \Psi_{\ell,j}$ \Comment{coflow to reject}

\KwRet $k^{\star}$
}
\end{algorithm}

There are two main phases in \wdcoflow. In the first phase, the algorithm works in iterations and in each iteration, it either accepts or rejects one coflow. The selected coflow is then removed from the current set of coflows $\cS$, which is initialized to $\cC$. In each iteration, \wdcoflow \ updates two vectors $\sigma$ and $\sigma^\star$ to keep track of candidate coflows:
$\sigma_{N-n+1}$ is set to the identity of the coflow selected in iteration $n=1,2,\ldots,N$, and $\sigma^{\star}_{N-n+1}$ is set to  the identity of the coflow rejected in that iteration, if any (otherwise, we set $\sigma^{\star}_{N-n+1} = 0$ and accept coflow $\sigma_{N-n+1}$).

In iteration $n$, \wdcoflow \  sweeps through the set of coflows $\cS$ to compute the total completion time $t_{\ell}=\sum_{k \in \cS} p_{\ell, k}$ of coflows in each port $\ell$. It then determines the bottleneck port $\ell_b$, i.e., the port $\ell$ with the largest completion time $t_{\ell}$. Let $k'$ be the coflow using port $\ell_b$ with the largest deadline. If $t_{\ell_b} \leq T_{k'}$, then coflow $k'$ can be scheduled as the last one on port $\ell_b$ and still satisfies its deadline. This coflow is therefore accepted by the algorithm and we set $\sigma_{N-n+1} = k'$ and $\sigma^{\star}_{N-n+1} = 0$. If on the contrary $t_{\ell_b} > T_{k'}$, this implies that at least one coflow among those using the bottleneck port will be late and therefore one of these coflows has to be rejected. \wdcoflowDP, a variant of \wdcoflow,  then uses a filtering algorithm described in Section \ref{subsec:filter} to compute a set $\cR \subseteq \cS_b$ of candidate coflows for rejection among those using the bottleneck port. For \wdcoflow, this filter is desactivated, so that $\cR = \cS_b$. The coflow $k^\star \in \cR$ to be rejected is then chosen as $k^{\star} = \arg\max_{k\in\mathcal{R}}\frac{1}{w_k}  \sum_{\ell \in \cL^\star} \Psi_{\ell,k}$, as explained in Section~\ref{subsec:coflow-rejection-rule}. \wdcoflow \  then  sets $\sigma_{N-n+1} = \sigma^{\star}_{N-n+1} = k^\star$.

The second phase of \wdcoflow \ is a post-processing phase intended at accepting unduly rejected coflows. Indeed, some coflows in $\sigma^{\star}$ could have been accepted if certain coflows that were rejected later would have been rejected earlier. To handle such cases, we use the function $\mathtt{RemoveLateCoflows}$ proposed in \cite{Dcoflow2022}. At the end of  the second phase of \wdcoflow, the estimated CCT of all coflows appearing in the order $\sigma$ is at most their deadline. 

 \lqt{
We revisit the example depicted in Fig.~\ref{fig:Example} to demonstrate the difference between \wdcoflow \ and  \csmha.  The execution of \wdcoflow\ on this example is presented in Table \ref{tab:Example}. In the initial step, \wdcoflow\ selects bottleneck ingress port $1$, which is used by coflows $k_{1}$ and $k_{2}$. It then calculates $\overline{\Psi}_{ k} = \sum_{\ell:\Psi_{\ell, k} < 0 } \Psi_{\ell, k}$ for both coflows  and  chooses the coflow that yields the largest $\bar\Psi_{ k}$ (in this case, $k_{1}$) to be scheduled last.
Since the remaining unscheduled coflows do not share any ports in the fabric, the specific ordering of these coflows does not impact the average CAR.
Given the final scheduling order, \wdcoflow\ yields a CAR of $\frac{4}{5}$, which is the optimal result, and is better than the average CAR of $\frac{1}{5}$ yielded by \csmha. In a general setting with $M$ machines, the CAR obtained using \csmha\ and \wdcoflow\ are respectively $\frac{1}{M}$ and $\frac{M-1}{M}$. 
}

\begin{table}[t]
	\caption{Execution of \wdcoflow  \ on the example of Fig.~\ref{fig:Example}. \label{tab:Example}}
	\centering
	\begin{tabular}{
			>{\raggedright}m{0.38\columnwidth}
			>{\centering}p{0.05\columnwidth}
			>{\centering}p{0.37\columnwidth}
		}
		\toprule 
		Unscheduled coflows (set $\mathcal{S}$) & $\ell_{b}$ & 
		$ \left\{ 
		\overline{\Psi}_{1},
		\overline{\Psi}_{2},
		\overline{\Psi}_{3},
		\overline{\Psi}_{4},
		\overline{\Psi}_{5}
		\right\} $ \tabularnewline
		\cmidrule[0.4pt](lr{0.12em}){1-1}%
		\cmidrule[0.4pt](lr{0.12em}){2-2}%
		\cmidrule[0.4pt](lr{0.12em}){3-3}%
		$\mathcal{S}=\left\{ \boldsymbol{k_{1}},\boldsymbol{k_{2}},k_{3},k_{4},k_{5}\right\} $ & 1 & $\left\{ -4\left(1+\varepsilon\right),-\varepsilon,\,\cdot\,,\,\cdot\,,\,\cdot\,\right\} $\tabularnewline
		$\mathcal{S}=\left\{ \boldsymbol{k_{2}},k_{3},k_{4},k_{5}\right\} $ & 1 & $\left\{ \,\cdot\,,0,\,\cdot\,,\,\cdot\,,\,\cdot\,\right\} $\tabularnewline
		$\mathcal{S}=\left\{ \boldsymbol{k_{3}},k_{4},k_{5}\right\} $ & 2 & $\left\{ \,\cdot\,,\,\cdot\,,0,\,\cdot\,,\,\cdot\,\right\} $\tabularnewline
		$\mathcal{S}=\left\{ \boldsymbol{k_{4}},k_{5}\right\} $ & 3 & $\left\{ \,\cdot\,,\,\cdot\,,\,\cdot\,,0,\,\cdot\,\right\} $\tabularnewline
		$\mathcal{S}=\left\{ \boldsymbol{k_{5}}\right\} $ & 4 & $\left\{ \,\cdot\,,\,\cdot\,,\,\cdot\,,\,\cdot\,,0\right\} $\tabularnewline
		\bottomrule
	\end{tabular}
\end{table}

In the following, we consider three variants of \wdcoflow, namely \dcoflow\ for unweighted coflows and \wdcoflow\ and \wdcoflowDP\ for weighted coflows. The first variant, \dcoflow\ corresponds to Algorithm 1 in \cite{Dcoflow2022} and  therefore assumes that all coflow weights are equal. The second variant, \wdcoflow,  is similar to \dcoflow \ but uses coflow weights in the coflow rejection rule, as described in Section~\ref{subsec:coflow-rejection-rule}. Finally, the third variant, \wdcoflowDP, works as \wdcoflow\  but uses a Dynamic Programming (DP) algorithm which plays the role of a filter that  restricts the choice of coflows that can be rejected. \wdcoflowDP\ is described in Section \ref{subsec:filter} below.


\subsection{Filtering Algorithm in \wdcoflowDP \label{subsec:filter}}
The coflow rejection rule discussed in Section~\ref{subsec:coflow-rejection-rule} is not necessarily optimal, even in the simple case of a single  port\footnote{If there is only one input port and one output port, the problem reduces to scheduling coflows on the minimum-capacity port.}. It turns out that, in this simple case, finding a maximum-weight feasible set of coflows is equivalent to the well-known scheduling problem of minimizing the weighted number of late jobs on a single machine, \lqt{a problem usually referred to as\footnote{\lqt{This  follows the notable triple $\alpha \vert \beta \vert \gamma$ notation proposed in \cite{graham1979optimization}, where $\alpha$ is the number of machines, $\beta$ is an optional list of job characteristics (not present in this case), and $\gamma$ is the objective function.}} $1 \Vert \sum w_j U_j$.
}
As it includes the ordinary knapsack problem as a special case, this problem is \textit{NP}-hard. Nevertheless, it can be solved by a dynamic programming algorithm within a pseudo-polynomial time bound of $\cO(nW)$, where $W=\sum_j w_j$, as we now explain~\cite{Lenstra1976}.

Without loss of generality, we assume that coflows are numbered in the EDD order, i.e., $T_1 \leq T_2 \leq \ldots, T_N$. As we assume that there is a single port $\ell$, we denote the processing time of coflow $k$ by $p_k$ instead of $p_{\ell, k}$. Let $P^{(j)}(w)$ denote the minimum total processing time for any feasible subset of coflows $1,\ldots,j$ that has total weight $w$. Initially, $P^{(0)}(0)=0$ and $P^{(0)}(w)=+\infty$ for all $w \in \{1,2,\ldots,W\}$. In the subsequent $n$ iterations $j=1,2,\ldots,n$, the variables $P^{(j)}(w)$ are computed as follows


\lqt{
\begin{equation} \label{eq:dp-iter}
P^{(j)}(w)=\begin{cases}
\min\left\{ P^{(j-1)}(w),\ P^{(j-1)}(w-w_{j})+p_{j}\right\} ,\\
\qquad\text{if }\ensuremath{P^{(j-1)}(w-w_{j})+p_{j}\leq T_{j}},\\
\ensuremath{P^{(j-1)}(w)},\text{ otherwise}.
\end{cases}
\end{equation}
}
At the end of the algorithm, the maximum weight of a feasible set is the largest value of $w$ such that $P^{(n)}(w)$ is finite. The maximum-weight feasible set is easily obtained with standard backtracking techniques. Interestingly, we note that when coflows have equal weights, or more generally when their processing times and weights are oppositely ordered, the problem $1 \Vert \sum w_j U_j$ can be solved in $\cO(n \log n)$ time with the Moore-Hogdson algorithm~\cite{Moore1968}.

The above DP algorithm is used by \wdcoflowDP \ in function $\mathtt{RejectCoflow}$ to compute the set $\cR$ of candidate coflows for rejection among those using the bottleneck port. The main advantage is that, as is easily proven,  \wdcoflowDP \  is optimal when there is only one input port and one output port. This is not the case of \dcoflow \ and \wdcoflow, even for coflows with equal weights. However, the downside is that the running time of  \wdcoflowDP \  is only pseudo-polynomial in the sum of coflow weights, whereas the complexity of  \wdcoflow \ is  easily proven to be $\cO(N^2)$ \cite{Dcoflow2022}. 


\subsection{Online Algorithm}
\label{subsec:dcoflow-online}

\lqt{
The three variants of \wdcoflow\ can also be performed in an online setting where coflows arrive sequentially and possibly in batches. 
For this, we introduce the update frequency, denoted as $f$.  This frequency represents the instances at which  \wdcoflow\ recomputes the coflow scheduling order. The updates can occur either when new coflows arrive (in which case $f$ is set to infinity) or periodically with a period of $1/f$. 
In the online scenario, the scheduler is aware of the flow volumes of the coflows currently in the system. However, it does not have knowledge of the volumes or release times of future coflows.

During each update instant, the scheduler recalculates a new order of the current coflows in the network. These include coflows that were scheduled in the previous update but have not yet completed, the ones that were rejected in the previous update but still have remaining time before their deadline, and the ones that have arrived during the update interval. The new ordering is determined based on the remaining volumes of the flows, rather than the original volumes. Note that coflows can be preempted in this process \cite{ChowdhuryThesis2015}. This recomputation of the schedule occurs at each update instant.
}


\section{Performance Evaluation 
\label{sec:Evaluation}}


\lqt{
In this section, we conduct an evaluation of our algorithms in comparison to state-of-the-art algorithms proposed in the literature. To ensure fairness and clarity, we begin by assigning equal weights to all coflows. This allows us to compare our algorithms against others that were developed without the ability to handle different coflow weights.
In the second part of this section, we extend our evaluation to consider the case with different coflow weights. 
}


\subsection{Simulation Setup
\label{subsec:Simulation-Setup}}

\lqt{
We evaluate via simulations\footnote{
The flow-level simulator and the implementation of all algorithms can be found at https://github.com/luuquangtrung/CoflowSimulator.} 
our proposed heuristics (three variants:   \dcoflow, \wdcoflow \ and \wdcoflowDP) along with some existing algorithms such as \csmha\footnote{
Only the centralized algorithm (\csmha) presented in \cite{Luo2016} is reimplemented, as it has been shown, in the same paper, to be better than the decentralized version ($\mathtt{D\textsuperscript{2}\text{-}CAS}$) in terms of CAR.
} 
and the solution provided by the optimization method \cdslp\  proposed in \cite{Tseng2019}. 
The relaxed version of \cdslp, named \cdslpa, is also implemented\footnote{
It is worth noting that both \cdslp\ and \cdslpa\ use the same decision variables $\{z_k\}_{k \in \cC}$ as those introduced in Problem~\eqref{prob:WCAR}. In \cdslp, $z_k$ are binaries, whereas in \cdslpa, $z_k$ are continuous numbers and can take values in the range $[0,1]$. For any solution obtained using \cdslpa, only coflows $k$ for which the corresponding ${z_k}$ strictly equals $1$ are considered as accepted ones.
}. By using the solution obtained from \cdslp\ as an upper bound, we can gain insight into how closely the evaluated algorithms approach the optimal solution. 
A concise overview of the reference algorithms has been provided in Sec.~\ref{sec:Introduction}. 
Furthermore, we conduct a comparative analysis by comparing our schedulers against two established algorithms, namely \sincronia\ \cite{Agarwal2018} and \varys\  \cite{Chowdhury2014} that aim to minimize the average CCT.
}


\lqt{
After obtaining the $\sigma$-order, the actual coflow resource allocation for our solution is performed using the greedy rate allocation algorithm $\mathtt{GreedyFlowScheduling}$ introduced in \cite{Agarwal2018}.  This algorithm reserves the entire bandwidth of a port for one flow at a time. It follows the order specified by $\sigma$, taking into account the corresponding coflow to which each flow belongs \cite{Agarwal2018}. Note that for \cdslp, \cdslpa, and \varys, the rate allocation is incorporated within the algorithm itself.
}


\lqt{
The network comprises $M$ machines connected to a non-blocking Big-Switch fabric, where each access port has a normalized capacity of  $1$. We assess the algorithms on small-scale and large-scale networks denoted as $[M, N]$, representing the fabric size and number of coflows ($N$) in the simulations. Small-scale networks consist of $M = 10$ machines, while large-scale networks have either $50$ or $100$  machines. Coflows in these networks are generated using either synthetic or real traffic traces.
}


\lqt{
\cdslp\ and \cdslpa\ are solved using the MILP solver $\mathtt{gurobi}$. Due to their high complexity, we only evaluate them on small-scale networks. The subsequent sections provide a comprehensive overview of the experimental setup, comparison metrics, and simulation results.
}

\lqt{
\vspace{0.2cm}
\noindent{\em Synthetic Traffic Traces.}
The synthetic traffic consists of two coflow types. Type-$1$ coflows have only one flow, whereas Type-$2$ coflows have a varying number of flows following a uniform distribution in $[2M/3, M]$. Each generated coflow is randomly assigned to either Class $1$ or Class $2$ with probability of respectively $0.6$ and $0.4$. 

Additionally, each coflow $k$ is assigned a random deadline within $[\text{CCT}_k^0, \alpha \text{CCT}_k^0]$, where $\text{CCT}_k^0$ represents the CCT of coflow $k$ \textit{in isolation}, and $\alpha$ is a positive real value in $[2, 4]$. A higher value of $\alpha$ indicates that the scheduler has more flexibility in meeting the coflow deadlines.
}

\lqt{
\vspace{0.2cm}
\noindent{\em Real Traffic Traces}
Real traffic traces are obtained from the Facebook dataset \cite{Chowdhury2014}. This dataset is based on a MapReduce shuffle trace collected from one of Facebook's $3000$-machine cluster with $150$ racks. The data traces contains a total of $526$ coflows with varying widths, ranging from small ones with only one flow to the largest ones with $21170$ flows. Detailed statistics of the Facebook dataset can be found in \cite{ChowdhuryThesis2015}. 

For each configuration $[M,N]$, $N$ coflows are randomly sampled from the Facebook dataset. They are only chosen from the coflows that have at most $M$ flows. The volume of each flow is already given by the dataset. 
}

\vspace{0.2cm}
\noindent{\em Weight Classes.} 
For \wdcoflow\ and \wdcoflowDP, we categorize coflows into two  classes of weights for both synthetic and real traffic. The weight assigned to each coflow reflects its importance level. Class-$1$ coflows are assigned a weight $w_1 = 1$, while Class-$2$ coflows are assigned a weight $w_2$ of either $2$ or $10$. The probability that a generated coflow falls into Class $1$ and Class $2$ are respectively $p_1$ and $p_2 = 1-p_1$.

\vspace{0.2cm}
\noindent{\em Metric.}
We evaluate the algorithms based on  the average weighted CAR, $\text{WCAR} = \frac{\sum_{k \in \cC} w_k z_k}{\sum_{k \in \cC} w_k}$ for the weighted setting. In the unweighted setting, WCAR is just the average CAR, where $w_k =  1$, $\forall k \in \cC$.   We also present  the gains in percentiles of each algorithm with respect to the solution provided by $\mathtt{CDS\text{-}LP}$ in terms of WCAR. These gains are calculated using the formula: $\text{average gain in WCAR} = \frac{\text{compared WCAR}}{\text{WCAR under } \mathtt{CDS\text{-}LP}} - 1.$

In addition, the per-class CAR is evaluated. For class $i\in\{1,2\}$, it is defined as the number of admitted coflows of class $i$ divided by the total number $N_i$ of coflows of this class, i.e.,  $\left ( \sum_{k \in \cC} \mathds{1}_{k,i} z_k \right )/N_i$, where $\mathds{1}_{k,i}=1$ if coflow $k$ is of class $i$, and $\mathds{1}_{k,i}=0$, otherwise.





\subsection{Scheduling Unweighted Coflows}
\label{subsec:scheduling-unweighted}


In this section, we assess the performance of our unweighted algorithm, \dcoflow, and compare it with other algorithms described in section~\ref{subsec:Simulation-Setup} using the same weights for all coflows. We recall that the unweighted case represents the evaluation of \dcoflow\ initially developed in \cite{Dcoflow2022}.

\subsubsection{Results with Offline Setting\label{sec:Eva:Offline}}

In the offline setting, we assume that all coflows arrive simultaneously with a release time of zero. For each simulation with a specific scale of the network and either synthetic or real traffic traces, we randomly generate $100$ different instances and calculate the average performance of all algorithms over $100$ runs.

\paragraph{{Average CAR Under Synthetic Traffic}}\label{sec:Eva:Offline:Rd}

\begin{figure}[h]
\centering
\subfloat[{Synthetic traffic
traces on a small-scale network.\label{fig:Off:Rate:Random:Small} }]{
\includegraphics[width=.7\columnwidth]{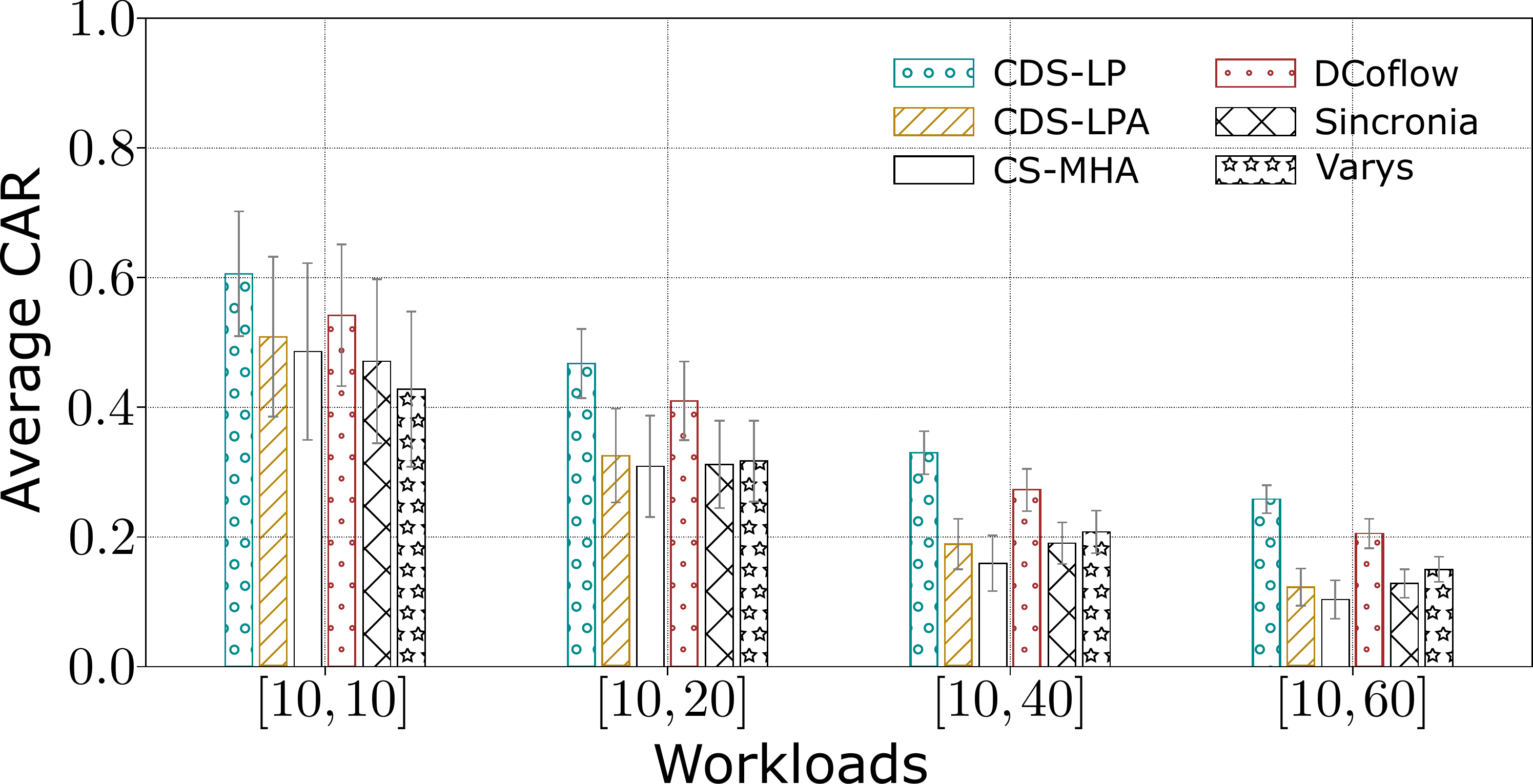}}

\vspace{-0.2cm}

\subfloat[{Synthetic traffic
traces on a large-scale network.\label{fig:Off:Rate:Random:Big} }]{
\includegraphics[width=.7\columnwidth]{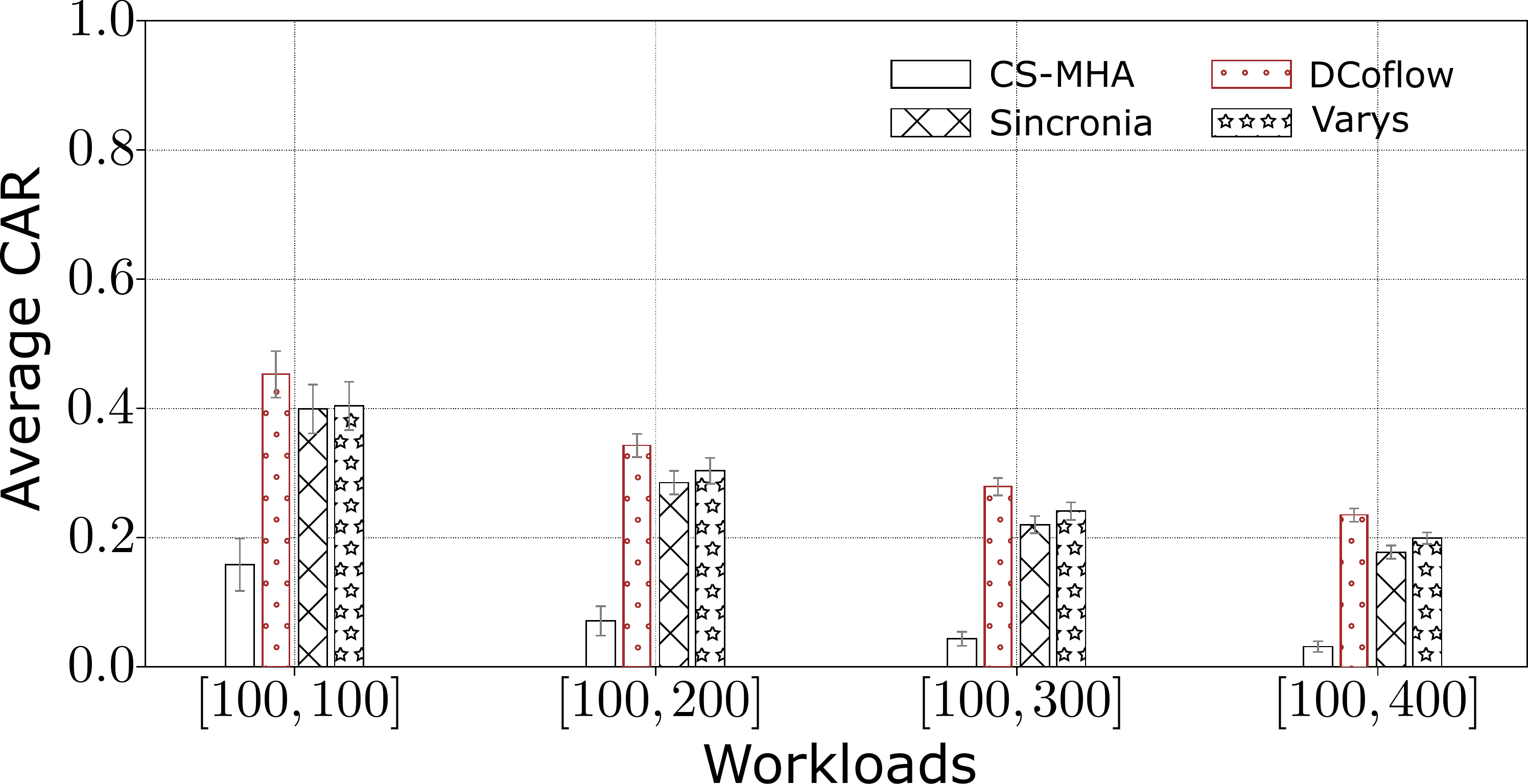}}
\caption{Average CAR with synthetic traffic traces using (a) small-scale and (b) large-scale networks. Each point in the x-axis represents the network $[M,N]$. \label{fig:Off:Rate:Random}}
\end{figure}

\begin{figure}[t]
\centering
\subfloat[{Synthetic traces. \label{fig:Off:Percentile:Random}}]{
\includegraphics[width=0.49\columnwidth]{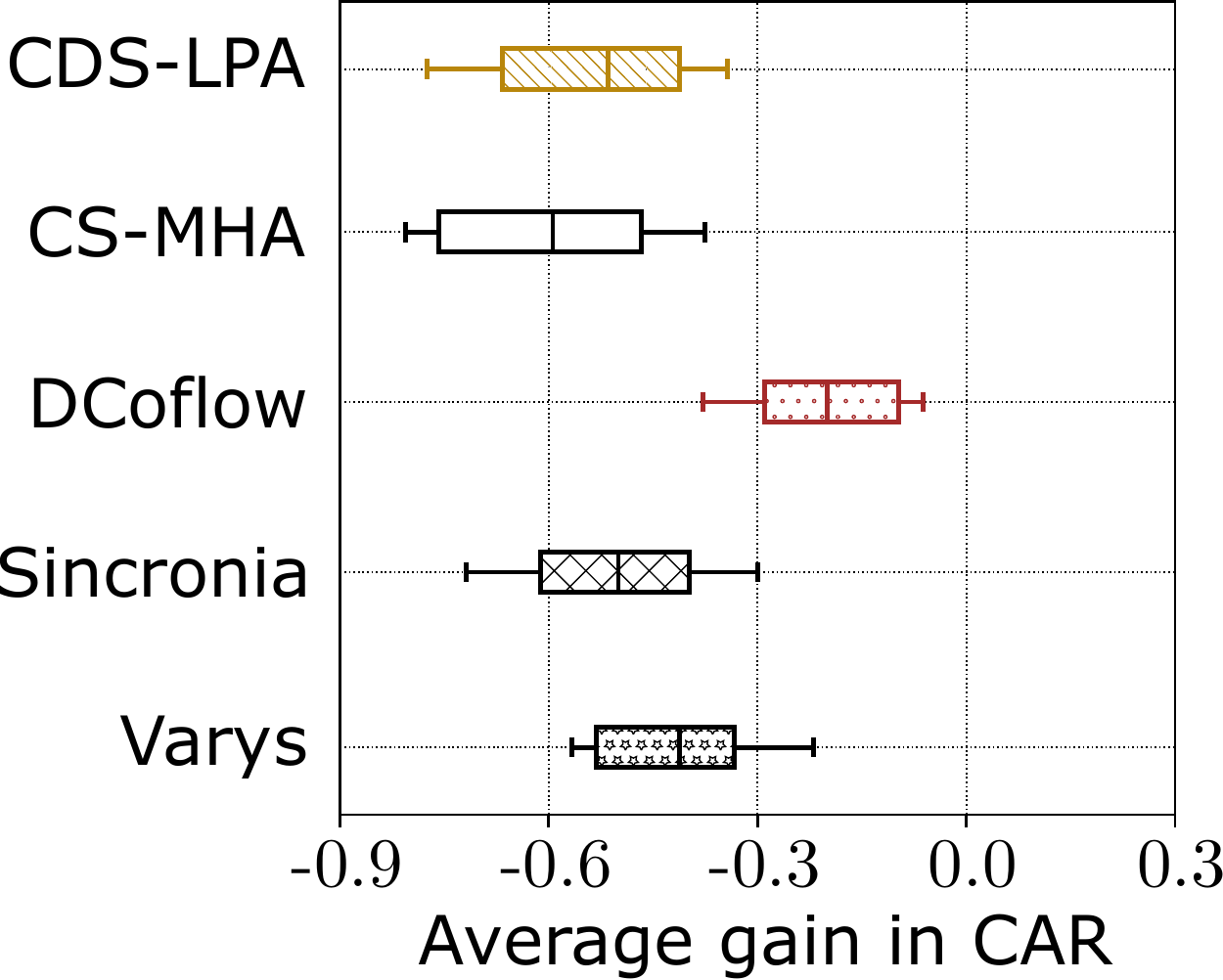}}
\subfloat[{Facebook traces.\label{fig:Off:Percentile:Facebook} }]{
\includegraphics[width=0.49\columnwidth]{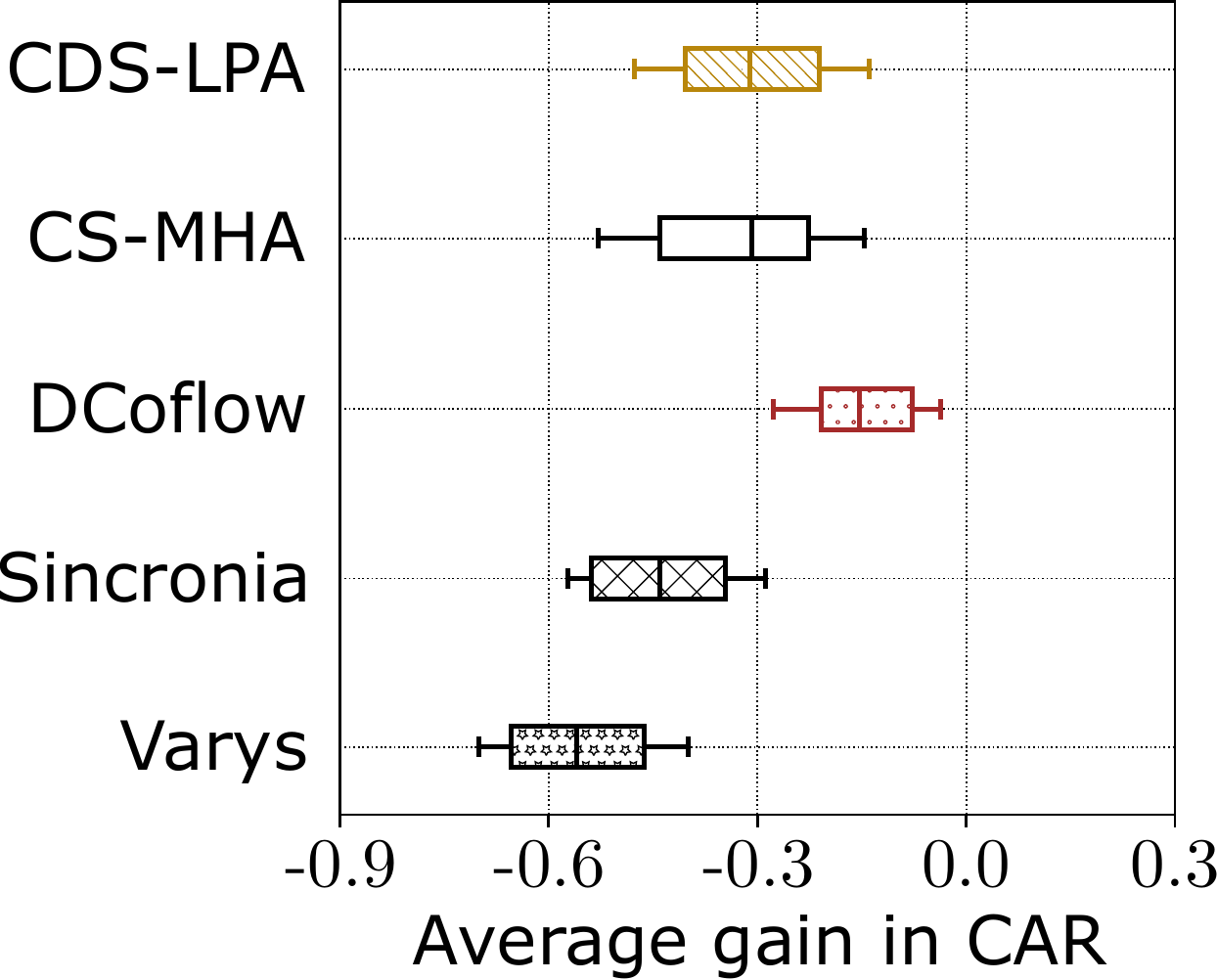}}
\caption{The $1$\textsuperscript{st}-$10$\textsuperscript{th} -$50$\textsuperscript{th}-$90$\textsuperscript{th}-$99$\textsuperscript{th} percentiles of the average gain in CAR with small-scale network $[10,60]$ using (a) synthetic and (b) Facebook traces. \label{fig:Off:Percentile}}
\vspace{-0.5cm}
\end{figure}

Figs.~\ref{fig:Off:Rate:Random:Small}--\ref{fig:Off:Rate:Random:Big} show the average CAR with respectively small-scale networks and large-scale networks. The percentile gains of each algorithm with respect to $\mathtt{CDS\text{-}LP}$  are shown in Fig.~\ref{fig:Off:Percentile:Random}, in terms of average CAR for the configuration $[10,60]$.

\lqt{
The results show that \dcoflow\ exhibits the closest performance to the optimal solution yielded by \cdslp\ in terms of CAR compared to all other algorithms. This holds true for both small- and large-scale networks. Surprisingly, \dcoflow\ even outperforms \cdslpa\, which is an approximation version of \cdslp\. These findings indicate the effectiveness of \dcoflow\ in achieving near-optimal performance for coflow scheduling.}
For instance, with the configuration $[10,10]$, \dcoflow\ improves the CAR on average by 
$6.5\%$,
$11.5\%$,
$15.1\%$, and
$26.6\%$,
compared respectively to 
\cdslpa, 
\csmha,
\sincronia, and
\varys. 
\lqt{
The improvement in average CAR becomes more pronounced as the load increases. Specifically,  the corresponding improvement on average CAR a configuration $[10,60]$ are
$67.2\%$,
$98.3\%$,
$59.9\%$, and
$36.8\%$ 
(see Fig.~\ref{fig:Off:Rate:Random:Small}). The improvement in performance is even more substantial when evaluated on a large-scale network. For example, compared to 
\csmha,
\sincronia, and
\varys, 
with the configuration $[100,400]$, the improvement in terms of average CAR are respectively
$648.1\%$,
$32.3\%$, and
$17.9\%$.
(see Fig.~\ref{fig:Off:Rate:Random:Big}). It is worth noticing how the performance of \csmha\ falls drastically when dealing with large-scale configurations.
This behavior can be attributed to the prioritization strategy of \csmha, which favors coflows that utilize a large number of ports over those that require only a few. 
In scenarios where there are numerous coflows with a small number of ports, the CAR of \csmha\ tends to approach zero (see detailed explanation of this behavior with a motivating example in Sec.~\ref{subsec:Motivating-Example}).
}



\lqt{
The results depicted in Figure~\ref{fig:Off:Percentile:Random} highlight that \dcoflow\ consistently achieves a smaller gap to the optimal solution across a wide range of percentile values compared to other algorithms. In particular, when compared to \sincronia,  \dcoflow\ improves the CAR in  $50\%$ of the  $100$ instances by $50\%$, and it achieves an approximately  $43\%$  improvement at the $99$\textsuperscript{th} percentile.
}

\paragraph{Average CAR Under Real Traffic Traces}\label{sec:Eva:Offline:Fb}

\begin{figure}[t]
\centering
\subfloat[{Facebook traffic traces on a small-scale network.\label{fig:Off:Rate:Facebook:Small} }]{
\includegraphics[width=.7\columnwidth]{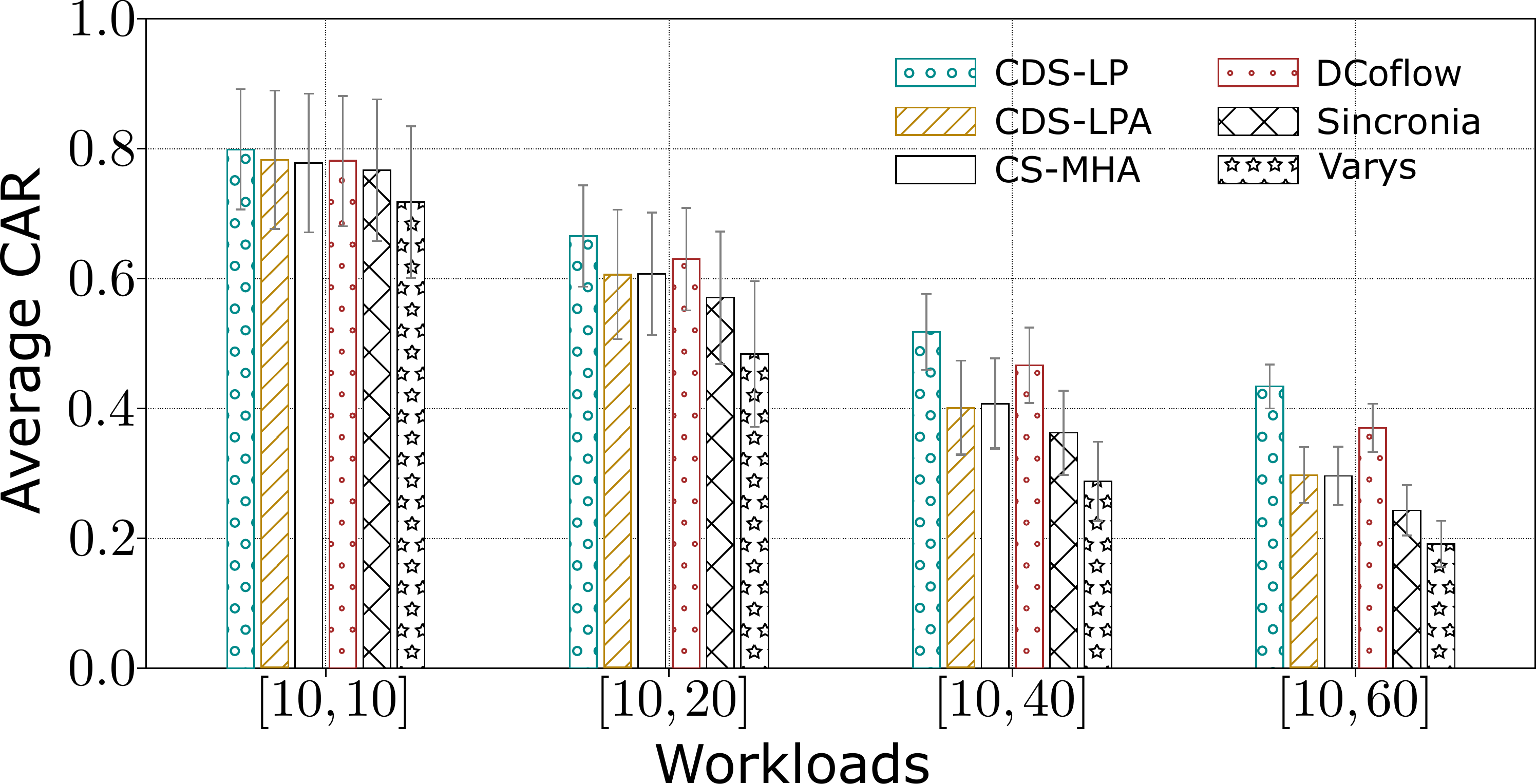}}

\vspace{-0.2cm}

\subfloat[{Facebook traffic traces on a large-scale network.\label{fig:Off:Rate:Facebook:Big} }]{
\includegraphics[width=.7\columnwidth]{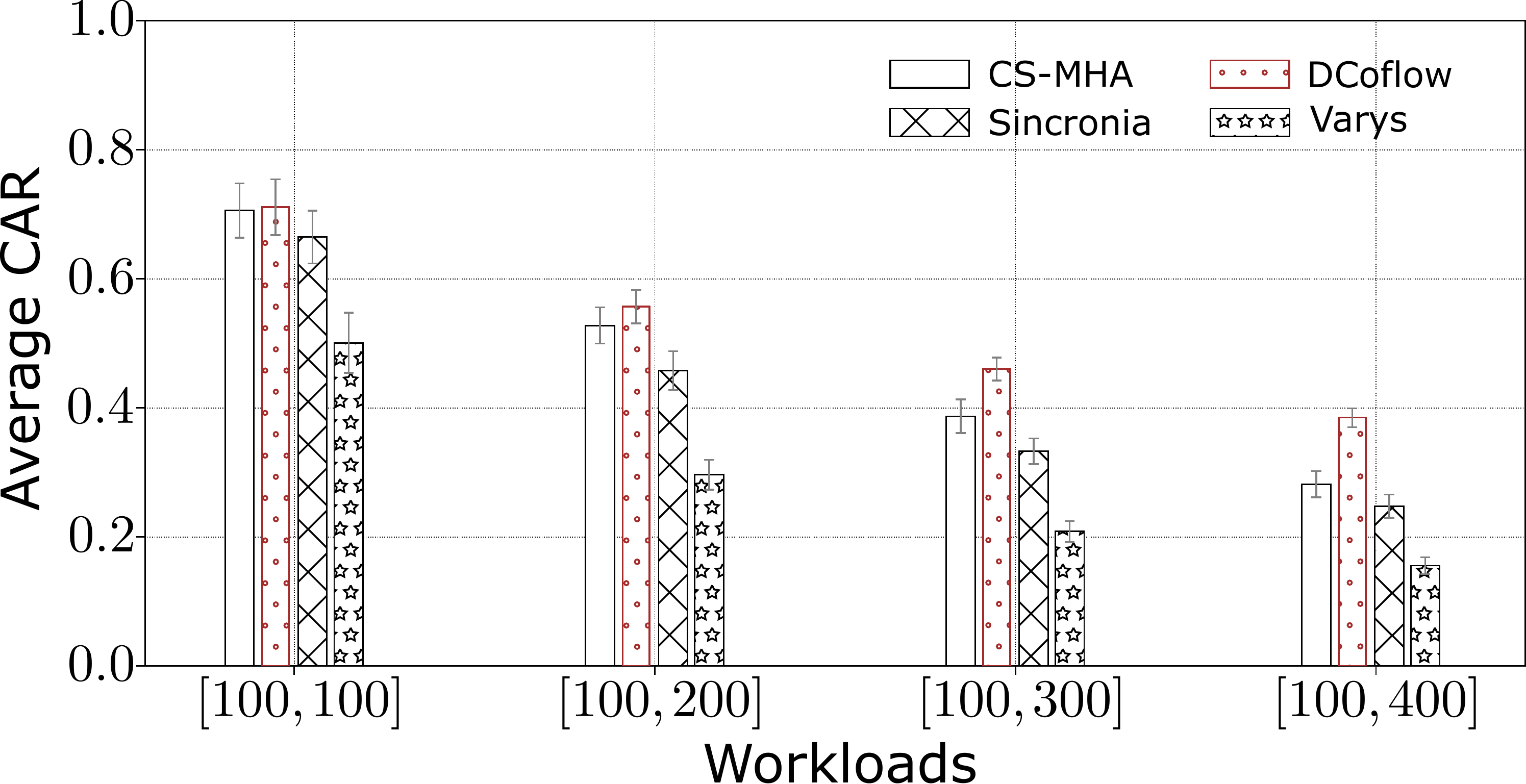}}
\caption{Average CAR with Facebook traces using (a) small-scale network and (b) large-scale network. Each point in the x-axis represents network $[M,N]$. \label{fig:Off:Rate:Facebook}}
\end{figure}

This section presents the results obtained with the Facebook traffic traces, using the same configurations as those used in Sec.~\ref{sec:Eva:Offline:Rd}. Figs.~\ref{fig:Off:Rate:Facebook:Small}--\ref{fig:Off:Rate:Facebook:Big} show the average CAR with respectively small- and large-scale networks. The gains in percentiles of each algorithm with respect to $\mathtt{CDS\text{-}LP}$, in terms of average CAR when using a $[10,60]$ network are shown in Fig.~\ref{fig:Off:Percentile:Facebook}. Similar to the results obtained using the synthetic traces (see Sec.~\ref{sec:Eva:Offline:Rd}), \dcoflow\ demonstrates a substantial improvement in terms of average CAR compared to other heuristics.
For instance, with a $[10,60]$ configuration, \dcoflow\  improves the average CAR by an average of 
$24.4\%$, 
$25\%$,
$52.2\%$, 
and $93.1\%$ 
compared respectively to 
\cdslpa, 
\csmha,
\sincronia, and 
\varys 
(see Fig.~\ref{fig:Off:Rate:Facebook:Small}). The improvement is even higher when performed on a large network configuration. For example, compared to \csmha, \sincronia, and \varys, on a $[100,400]$ network, the improvement in terms of average CAR are respectively 
$36.6\%$, 
$55.3\%$, and 
$147.5\%$.


\lqt{
Moreover, the results in Fig.~\ref{fig:Off:Percentile:Facebook} show that \dcoflow\ consistently achieves a smaller gap to the optimal solution across various percentiles compared to the other algorithms. Specifically, compared to \sincronia, \dcoflow\ improves the CAR in $57\%$ of $100$ instances by $50\%$, and it achieves around $35\%$ at the $99$\textsuperscript{th} percentile.
}

\paragraph{Prediction Error of $\mathtt{DCoflow}$}


\lqt{
It is worth noticing that the final solution provided by \dcoflow\ does not necessarily guarantee that every coflows in $\sigma$ will eventually meet their deadlines. The estimated CCT of coflows may differ from the actual CCTs obtained after the rate allocation process due to the coupling  between input and output ports. The prediction error of \dcoflow\ represents the gap between the estimated CAR and the actual CAR after resource allocation. This error is given by $(|\sigma| - |\hat{\sigma}|)/|\sigma|$, where $\hat{\sigma} \subseteq \sigma$ is the subset of coflows in $\sigma$ that meet their deadlines after applying the actual rate allocation using $\mathtt{GreedyFlowScheduling}$.

In the simulations presented in Sec.~\ref{subsec:scheduling-unweighted}, we observe an average CAR prediction error  of below $3.6\%$ of \dcoflow\  for both synthetic and real traffic traces.
}

\subsubsection{Online Setting\label{sec:Eva:Online}}


\lqt{
We now present  a series of numerical results regarding the performance of the online version of \dcoflow. The evaluation metric used is the average CAR obtained from $40$ instances. In each instance, coflows arrive sequentially based on a Poisson process with a rate of $\lambda$, i.e., the inter-arrival time of coflows is exponentially distributed with rate $\lambda$. By default, coflow priorities are computed when a new coflow arrives ($f=\infty$), unless otherwise specified.
}



\lqt{
We compare the average CAR achieved by \dcoflow\ with the online version of Varys with deadline \cite{Ma2016}, \csmha, and \sincronia. We examine the impact of two key parameters: (\textit{i}) the coflow arrival rate $\lambda$ and (\textit{ii}) the frequency $f$ at which coflow priorities are recomputed. 
}


\paragraph{Impact of Arrival Rate}


\lqt{
We begin by examining the impact of the arrival rate $\lambda$ on the CAR achieved by different algorithms. The CAR is averaged over $40$ instances, each consisting of $4000$ coflow arrivals. 
The deadline for each coflow $k$ is randomly selected from a uniform distribution in the range $[\text{CCT}_k^0, 4\text{CCT}_k^0]$. Two scenarios are considered: a small fabric with $M = 10$ machines and a large fabric with $M = 50$ machines. In each scenario, we present the results for the following values of $\lambda$: $\lambda = 8$, $\lambda = 12$, $\lambda = 16$, and $\lambda = 20$. 
}

\begin{figure}[t]
\centering
\subfloat[{Small fabric.\label{fig:On:Rate:Random:Rachid:M10} }]{
\includegraphics[width=.48\columnwidth]{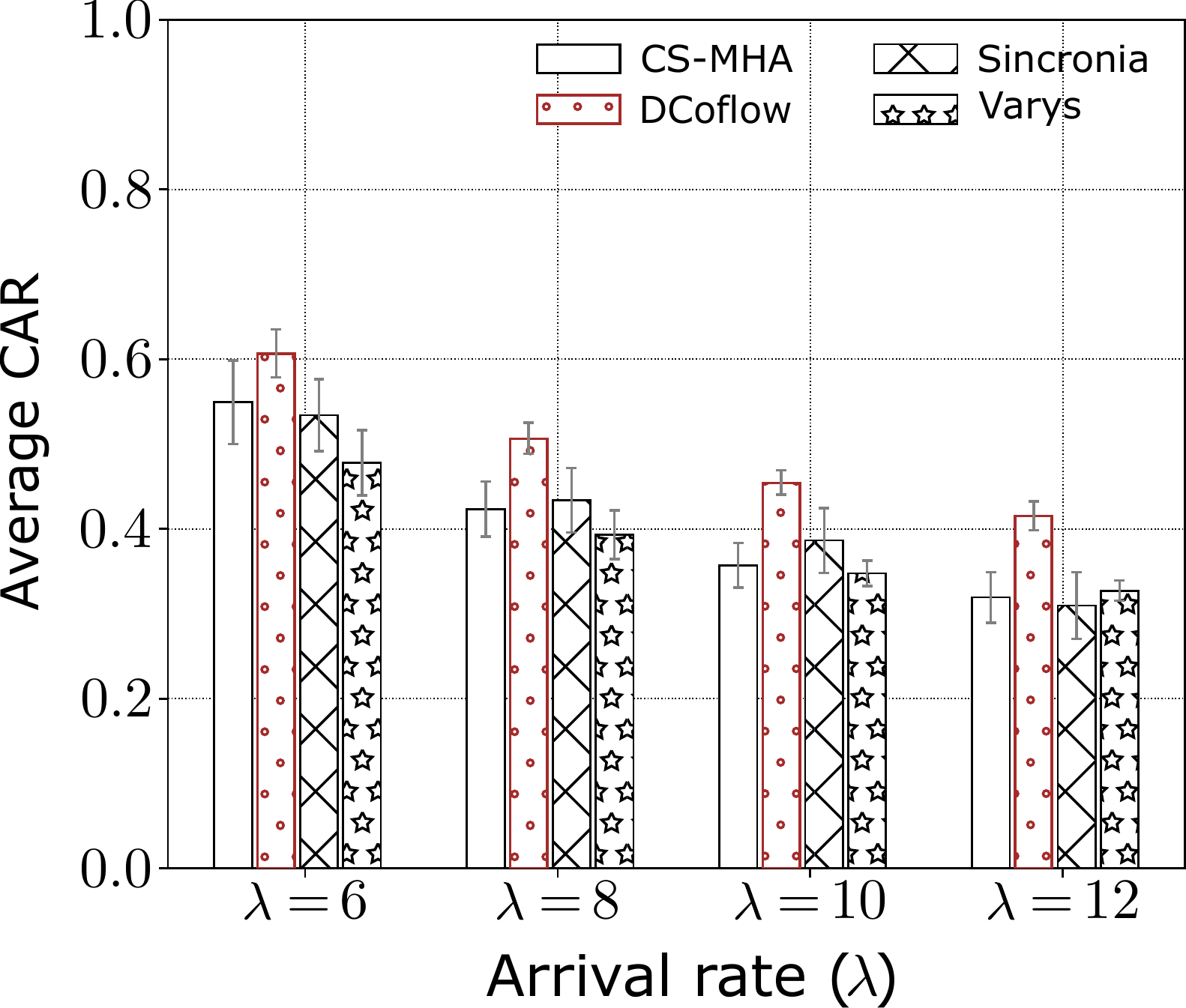}}
\subfloat[{Large fabric.\label{fig:On:Rate:Random:Rachid:M50} }]{
\includegraphics[width=.48\columnwidth]{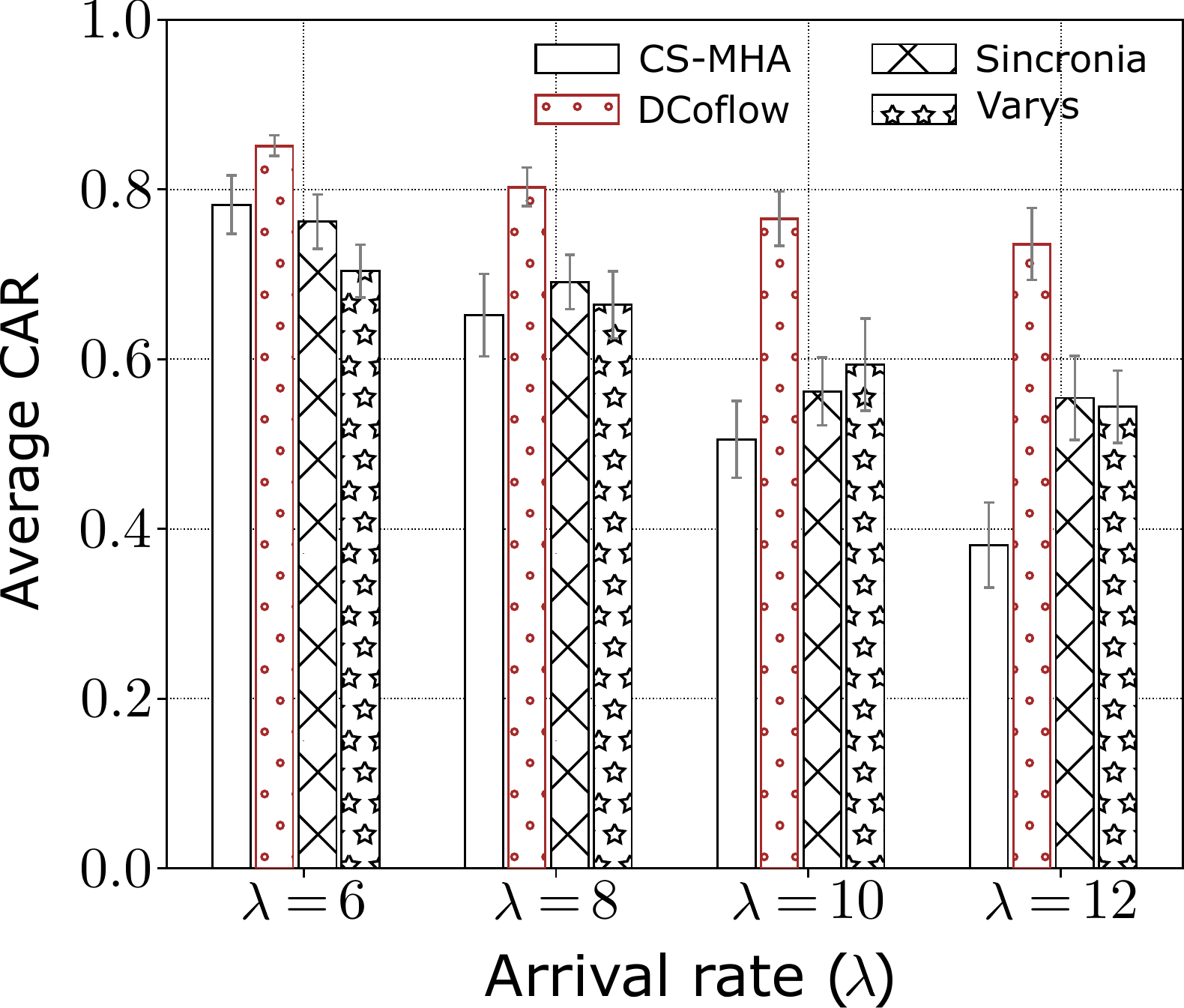}}
\caption{Average CAR using synthetic traffic with varying $\lambda$ and (a) $M = 10$ and (b) $M = 50$. \label{fig:On:Rate:Random:Rachid}}
\end{figure}

\lqt{
Figs.~\ref{fig:On:Rate:Random:Rachid:M10} and \ref{fig:On:Rate:Random:Rachid:M50} depict the results for respectively the small and large network. These results show that \dcoflow\ obtains a higher average CAR for all values of $\lambda$. Moreover,  the gain performance of  \dcoflow\  with respect to the other algorithms increases with the value of $\lambda$. While the other algorithms may exhibit similar CAR  in lightly loaded fabrics, \dcoflow\ clearly outperforms them when the network is heavily congested.
}

\begin{figure}[htb]
\centering
\subfloat[{Small fabric.\label{fig:On:Rate:Facebook:Rachid:M10C4000F60} }]{
\includegraphics[width=.48\columnwidth]{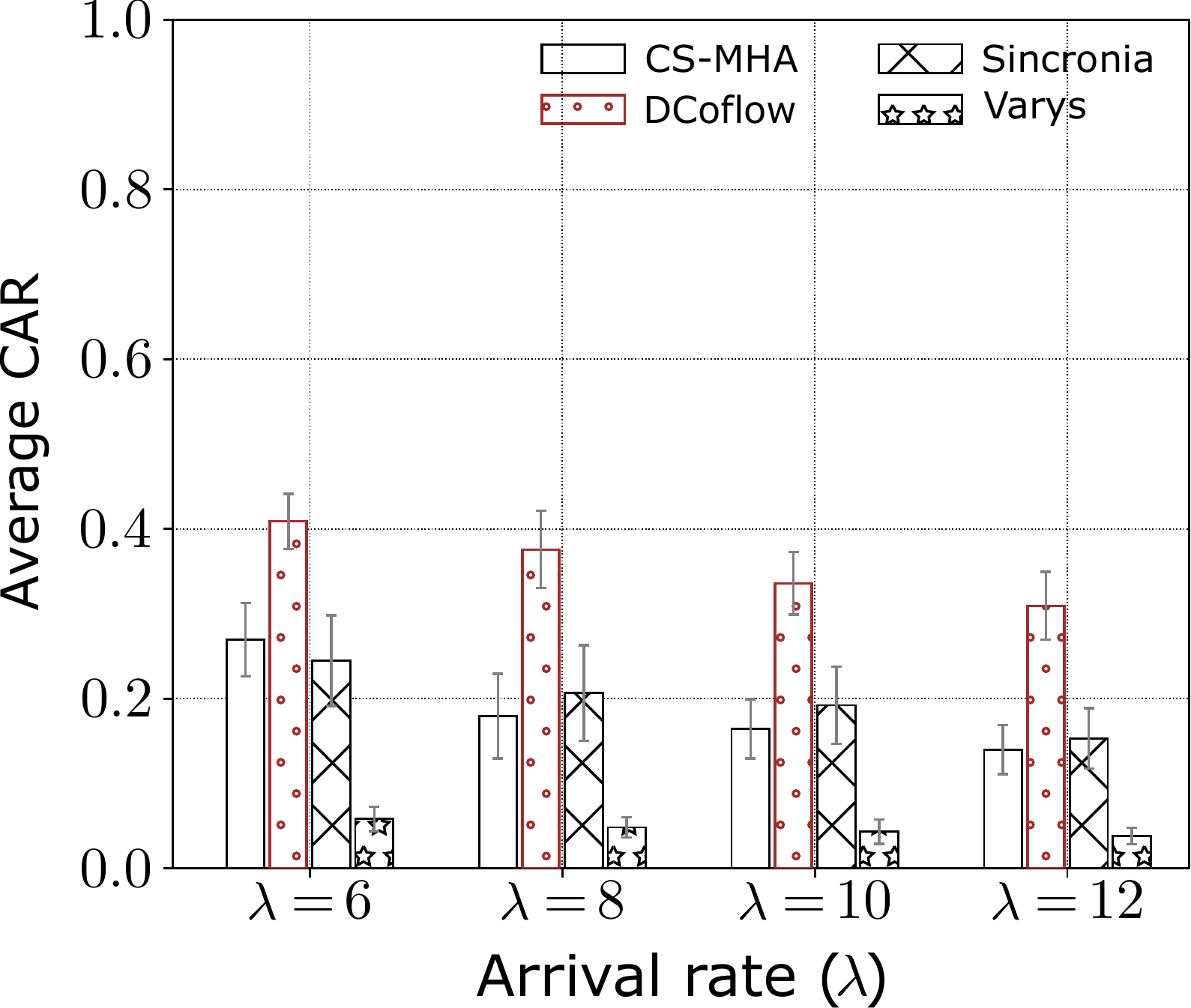}}
\subfloat[{Large fabric.\label{fig:On:Rate:Facebook:Rachid:M100C4000F200} }]{
\includegraphics[width=.48\columnwidth]{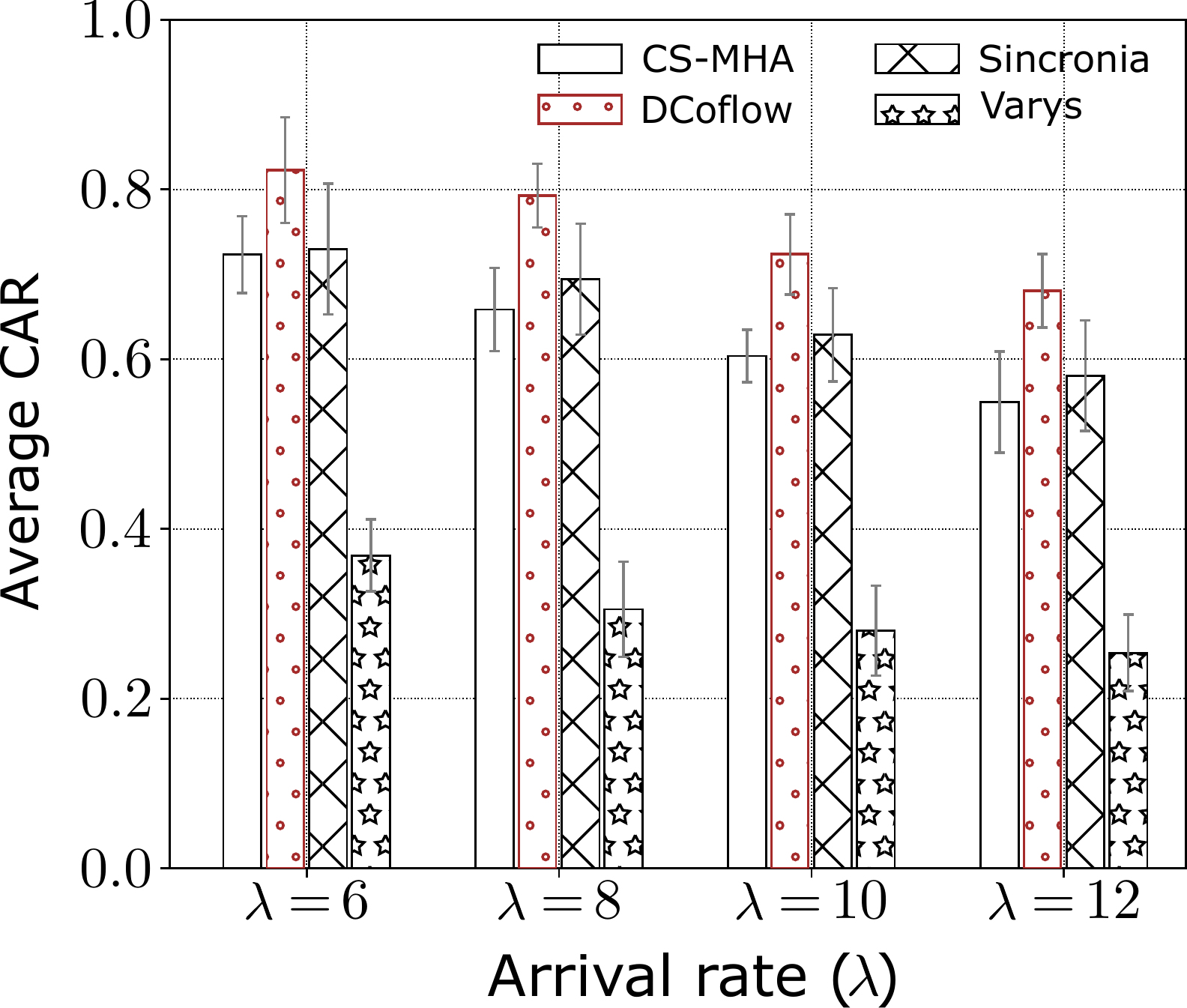}}
\caption{Average CAR using Facebook traffic with varying $\lambda$ and (a) $M = 10$ and (b) $M = 100$. \label{fig:On:Rate:Facebook:Rachid}}
\end{figure}

\lqt{
Figs.~\ref{fig:On:Rate:Facebook:Rachid:M10C4000F60} and \ref{fig:On:Rate:Facebook:Rachid:M100C4000F200} show respectively the average CAR when using the configuration of $M=10$ and $M=100$, both with $4000$ coflows, with the Facebook dataset. Similar to what was observed with the synthetic traffic traces, \dcoflow\ significantly outperforms all other methods. When dealing with a highly congested network (i.e., with $M=10$), again \dcoflow\ yields a higher gain compared to the other methods. For instance, when $M=100$ (see Fig.~\ref{fig:On:Rate:Facebook:Rachid:M100C4000F200}), \dcoflow\ achieves $9.3\%$ higher CAR than $\mathtt{Sincronia}$ , while with $M=10$, the gap becomes $16.4\%$ (see Fig.~\ref{fig:On:Rate:Facebook:Rachid:M10C4000F60}).
}

\paragraph{Impact of Update Frequency}


\lqt{
To evaluate the impact of the update frequency $f$ on the average CAR, the following values of $f$ are considered: $f=\frac{\lambda}{2}$, $f=\lambda$, $f=2\lambda$, and $f=\infty$. Recall that $f=\infty$ indicates that priorities are recomputed upon each arrival of a new coflow. We assume that $M=10$ and compute the CAR by averaging over $40$ instances. For each instance, $8,000$ coflow arrivals are generated, following a Poisson process of rate $\lambda$. The deadline of a coflow $k$ follows a uniform distribution in the range $[\text{CCT}_k^0, 2\text{CCT}_k^0]$. We examine the average CAR for different values of $f$ ($f \in \{ \frac{\lambda}{2}, \lambda, 2\lambda,  \infty \}$) and of the arrival rate which takes values in the range $[2, 10]$.
}



\begin{figure}[t]
\centering
\subfloat[{Without batch.\label{fig:On:Rate:Random:Olivier:nobatch} }]{
\includegraphics[width=.48\columnwidth]{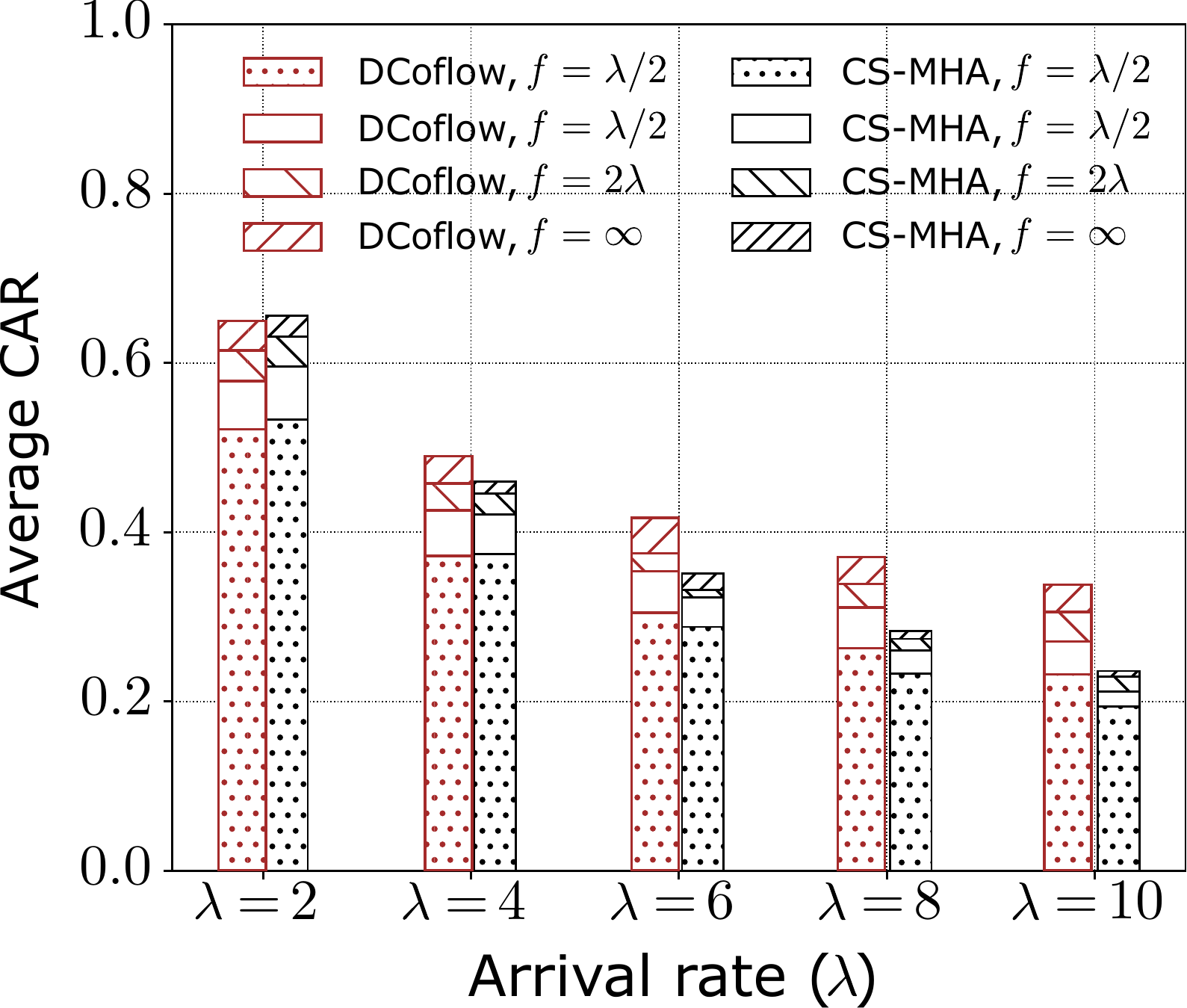}}
\subfloat[{With batch.\label{fig:On:Rate:Random:Olivier:batch} }]{
\includegraphics[width=.48\columnwidth]{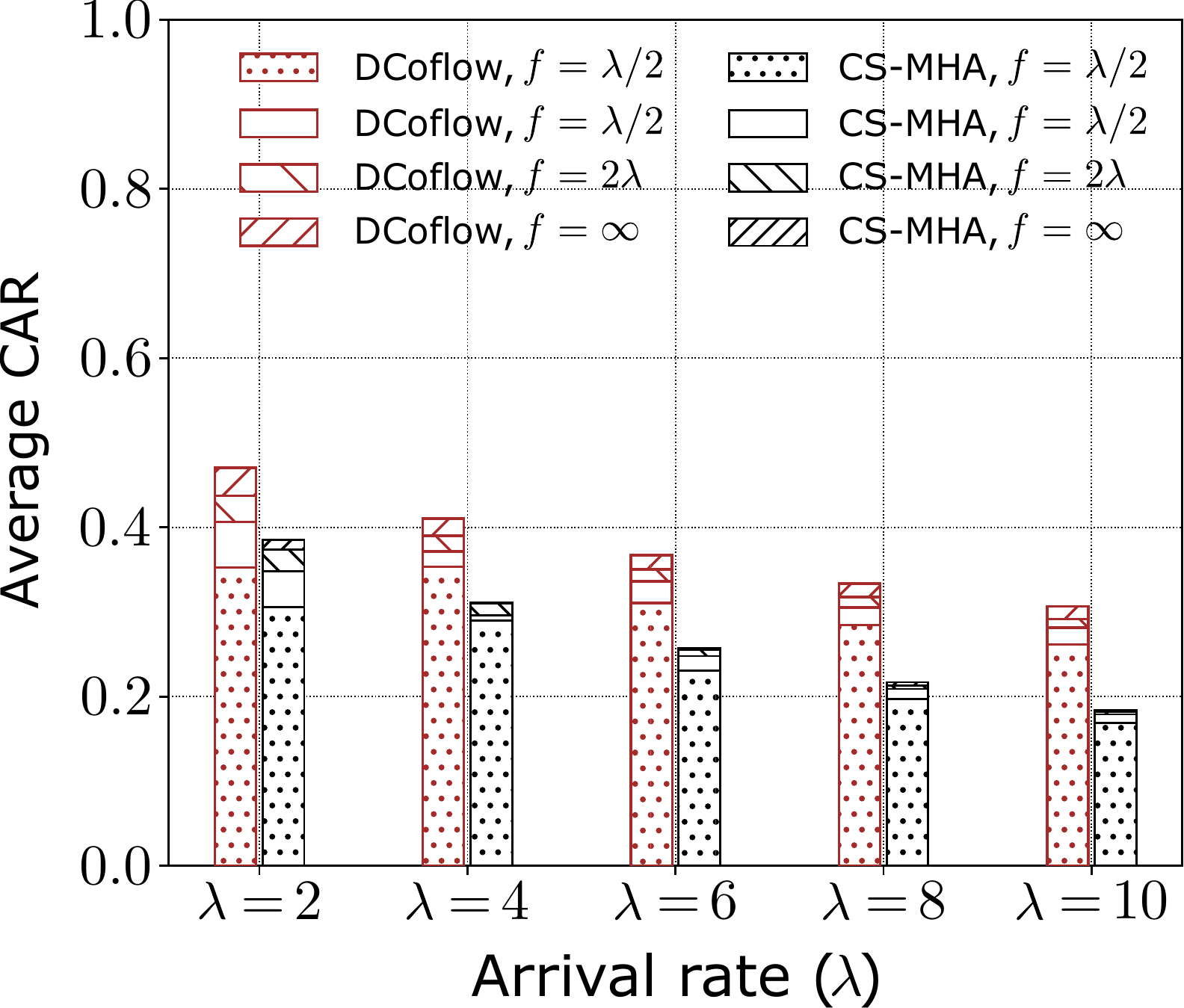}}
\caption{Average CAR of $\mathtt{DCoflow\_v1}$ and $\mathtt{CS\text{-}MHA}$ using synthetic traffic with $[10,8000]$ and varying $\lambda$, when obtaining (a) one single coflow per arrival; and (b) a random batch of coflow per arrival. \label{fig:On:Rate:Random:Olivier}}
\end{figure}

\lqt{
Fig.~\ref{fig:On:Rate:Random:Olivier:nobatch} shows the results obtained from a simulation, in which each arrival corresponds to one single coflow. 
Similar to previous findings, for a low arrival rate $\lambda$, both \dcoflow\ and \csmha\ achieve a similar average CAR performance: for $\lambda=2$,  \csmha\ achieves a slightly higher CAR than  \dcoflow). 
But when the network is highly congested, \dcoflow\ significantly outperforms  \csmha. 
Additionally, increasing the frequency $f$ has a noticeable positive impact on the CAR for both algorithms. 
For example, for $\lambda=2$ (resp. $\lambda=10$), updating coflow priorities upon each arrival (i.e., $f=\infty$) instead of using the periodic scheme with $f=\frac{\lambda}{2}$ leads to an average CAR increase of $52\%$ (resp. $46\%$). 
These results suggest that there is a trade-off between the computational complexity of updating coflow priorities at a high frequency and the  achieved CAR.
Fig.~\ref{fig:On:Rate:Random:Olivier:batch} shows a similar analysis, but this time we assume that coflows arrive in batches. The size of each batch is randomly drawn from a uniform distribution $\mathcal{U}([5,15])$. 
In this scenario, to ensure that the coflow arrival rates are comparable to the previous setting (Fig.~\ref{fig:On:Rate:Random:Olivier:nobatch}), where coflows arrive individually, we divide the batch arrival rate by 10. This adjustment allows us to maintain the same coflow arrival rates for both settings.

The results achieved for simulations with batch arrivals are similar to those obtained with the previous setting, but we note that \dcoflow\ continues to exhibit significant gains over \csmha. 
Additionally, we observe that the benefits of using a higher update frequency are relatively lower in this scenario. For instance, when $\lambda=10$, increasing the update frequency from $f=\frac{\lambda}{2}$ to $f=\infty$ results in only a $17\%$  increase in the average CAR.
}


\subsection{Scheduling Weighted Coflows}

We  now evaluate the weighted versions of our proposed algorithm, \wdcoflow\ and \wdcoflowDP, along with \csdp, \cdslp, and \cdslpa. \csdp\ is the adapted version to the weights of \csmha\ presented in \cite{Luo2016} (in which the Moore-Hogdson algorithm is replaced by the DP algorithm in Section \ref{subsec:filter}), whereas \cdslp\ and its relaxed variant \cdslpa  \ are straightforward adaptations of the linear-programming methods proposed in \cite{Tseng2019} to account for coflow weights. By using the solution derived from \cdslp\ as an upper bound, we can get the sense of how close the algorithms are to the optimum.

\subsubsection{Offline Setting\label{sec:Eva:Off}}
In the offline setting, we consider that all coflows arrive at the same time, i.e., their release time is zero. For each simulation with a specific scale of the network and either synthetic or real traffic traces, we randomly generate $100$ different instances and compute the average performance of algorithms over $100$ runs. We  evaluate  \wdcoflow\ and \wdcoflowDP\ against  existing algorithms for the offline case.

\paragraph{Synthetic Traffic\label{sec:Eva:Off:Syn}}

\begin{figure}[t]
\centering
\subfloat[{Small-scale networks.\label{fig:Off:Syn:WCAR:Small} }]
{
\includegraphics[width=4.4cm, height=3.6cm]{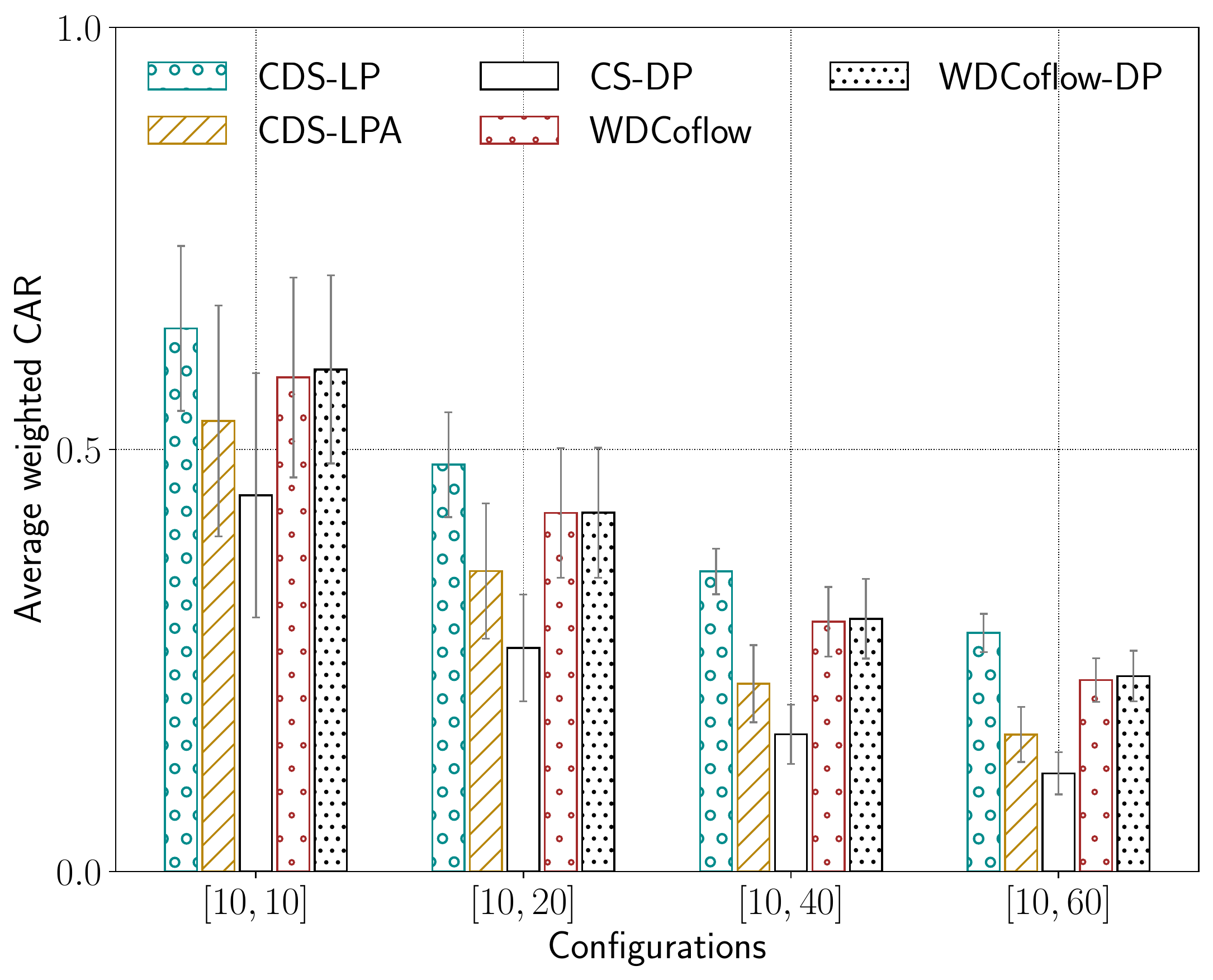}
}
\subfloat[{Large-scale networks.\label{fig:Off:Syn:WCAR:Big} }]{
\includegraphics[width=4.4cm, height=3.6cm]{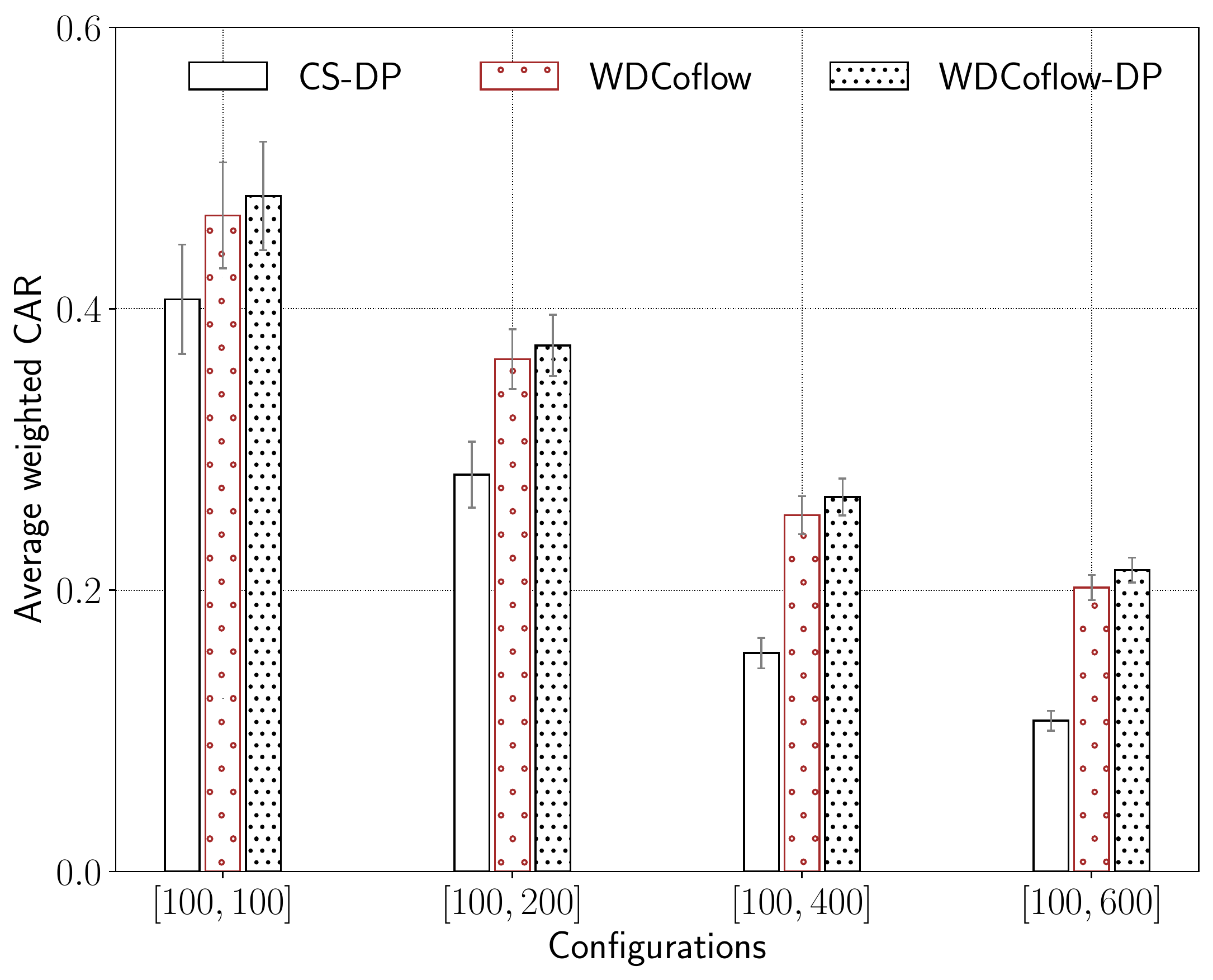}}
\caption{Average WCAR with coflows of all classes using synthetic traffic traces and (a) small-scale and (b) large-scale networks. The values of $(p_2,w_2)$ are set to $(0.2,2)$. Each point in the x-axis represents the network $[M,N]$. \label{fig:Off:Syn:WCAR}}
\end{figure}

Figs.~\ref{fig:Off:Syn:WCAR:Small} and \ref{fig:Off:Syn:WCAR:Big} shows the average WCAR with synthetic traffic traces using small-scale ($M = 10$) and large-scale ($M = 100$) networks.  It is observed that our proposed heuristics (\wdcoflow\ and \wdcoflowDP)  are closest in terms of WCAR to the optimum (\cdslp) than all other algorithms in both small- and large-scale network configurations. \wdcoflowDP\  yields a slightly higher performance compared to \wdcoflow\ (around $4.2\%$) when using large-scale network settings (see Fig.~\ref{fig:Off:Syn:WCAR:Big}).  For small scale network,  we observe that  the optimum \cdslp \ obtain only a  gain of $9\%$  and $7\%$ compared to   \wdcoflow\ and \wdcoflowDP\   respectively for $N=10$ .    On the other hand, CDS-LPA and CS-DP are far from the optimal solution by $17\%$ and $30\%$ respectively.  For the worst case when $N=60$,   \wdcoflow\ and \wdcoflowDP\   are far from the optimal solution by $19\%$ and $17\%$ respectively, but CDS-LPA and CS-DP moves away from the optimum by $42\%$ and $58\%$ respectively. 

Now, for large-scale networks, we observe that  \wdcoflow\ and \wdcoflowDP\  obtain a significant performance improvement compared to \csdp. Indeed, for $N=100$,  \wdcoflow\ and \wdcoflowDP\  obtain a gain of $14\%$ and $18\%$ respectively.  When the number of coflows increases, both algorithms obtain gains up to $184\%$ for \wdcoflow\  and $192\%$ for \wdcoflowDP\  compared to  \csdp.


\begin{figure}[t]
\centering
\subfloat[{Small-scale networks.\label{fig:Off:Syn:ClassCAR:Small} }]{
\includegraphics[width=4.4cm, height=3.6cm]{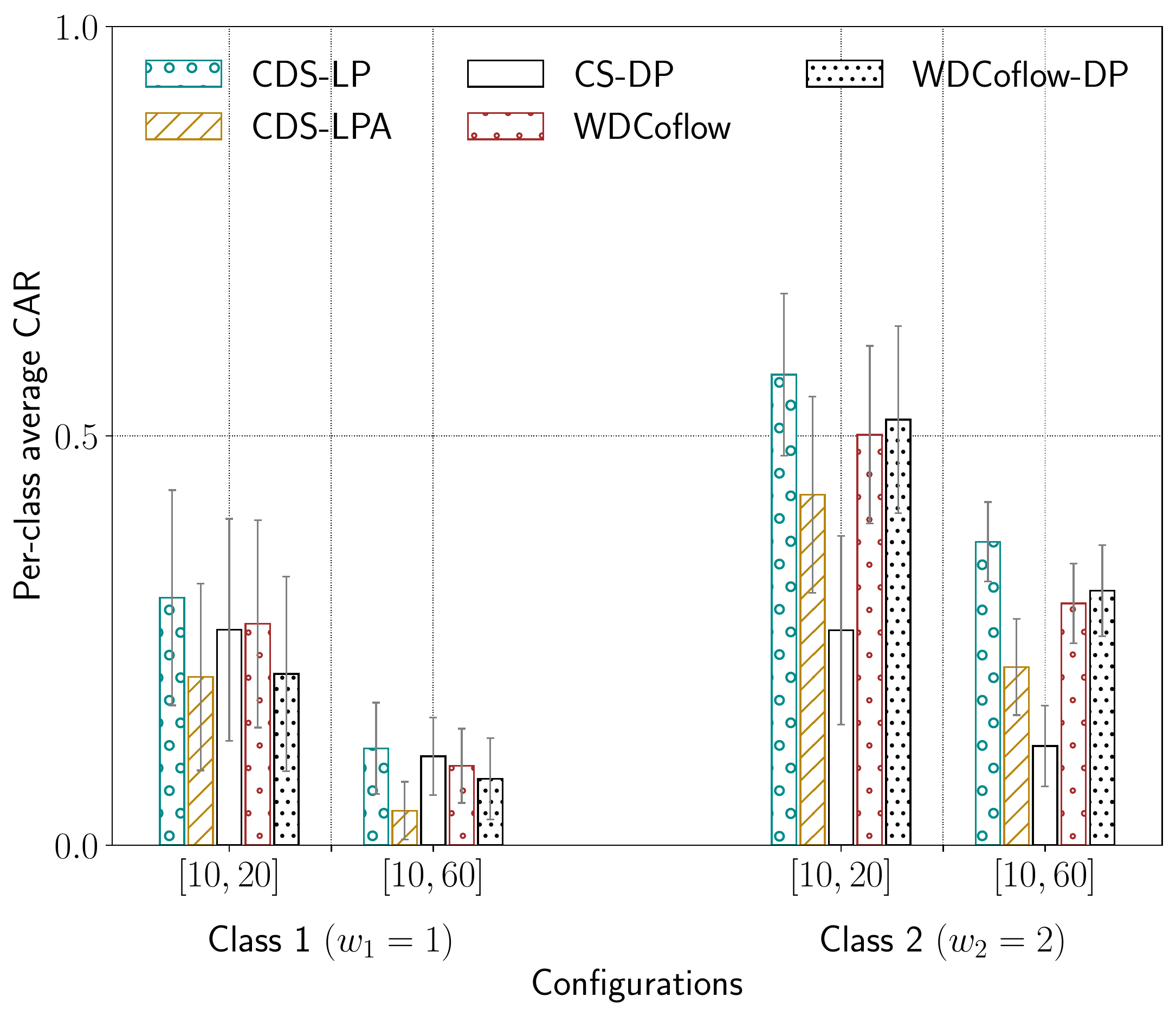}}
\subfloat[{Large-scale networks.\label{fig:Off:Syn:ClassCAR:Big} }]{
\includegraphics[width=4.4cm, height=3.6cm]{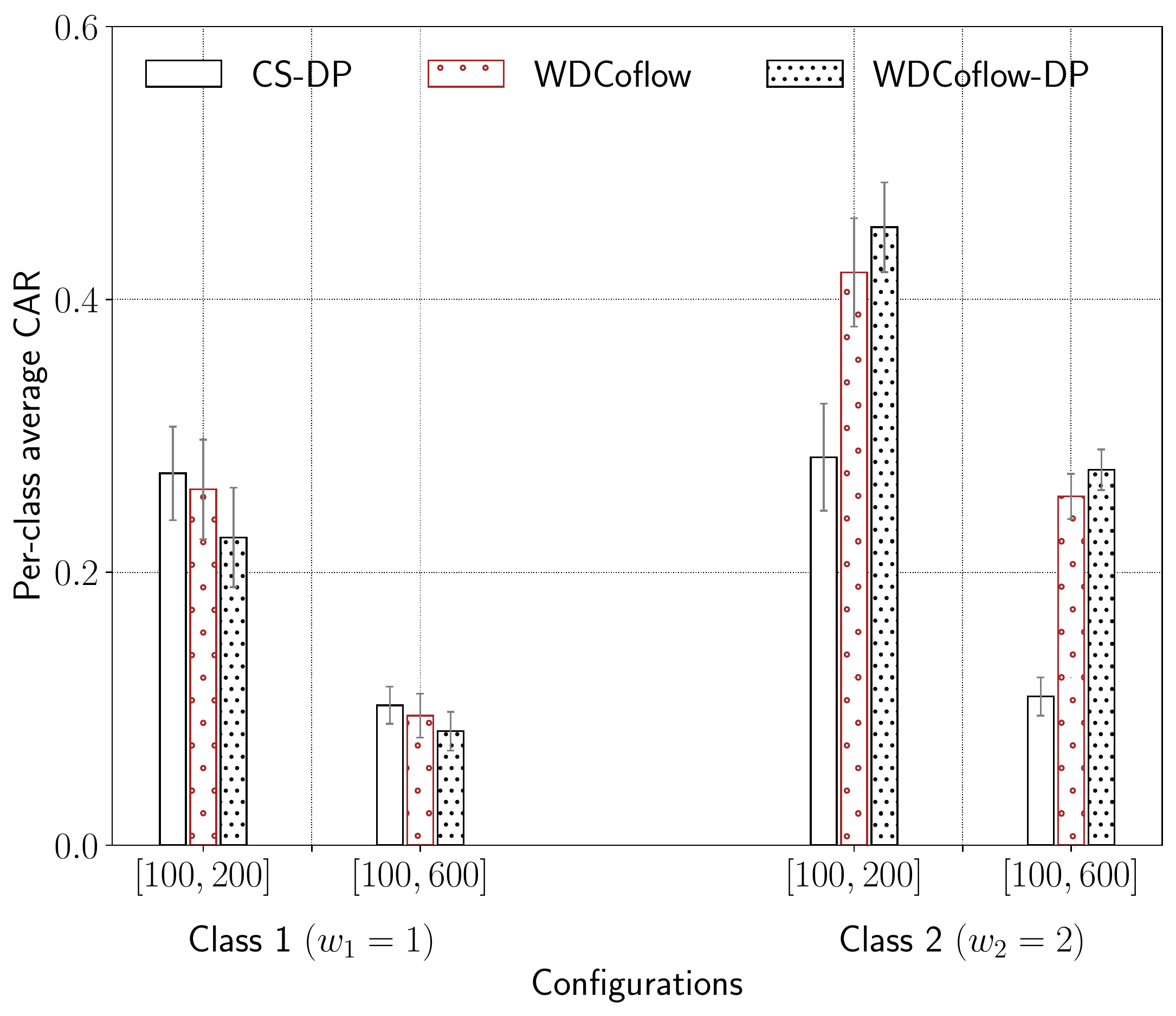}}
\caption{Average per-class CAR with synthetic traffic traces using (a) small-scale and (b) large-scale networks. The value of $p_2$ is set to $0.2$. Each point in the x-axis represents the network $[M,N]$. \label{fig:Off:Syn:ClassCAR}}
\end{figure}

With respect to the performance of each class,
Figs.~\ref{fig:Off:Syn:ClassCAR:Small} and \ref{fig:Off:Syn:ClassCAR:Big} show the average WCAR of each coflow class  using small and large-scale networks.   As expected, the performance is even more significant for traffic of Class $2$  since  \wdcoflow\ and \wdcoflowDP\ consider both the network conditions and coflows' importance  to perform the scheduling, while \csdp\  prioritizes coflows that use a large number of ports over those that use a few.  In Fig.~\ref{fig:Off:Syn:ClassCAR:Big}, we can see that  \wdcoflow\ and \wdcoflowDP\ perform about $21\%$ and $51\%$ for $N=100$ and $247\%$ and $258\%$ for $N=600$ better than \csdp\  for Class $2$. 
For class-$1$ coflows, our heuristic achieves a moderate gain of up to 10\% for $N=600$ compared to \csdp.

\begin{figure}[t]
\centering
\subfloat[{Varying $p_2$ with fixed $w_2$.\label{fig:Off:Syn:ChangeParam:p2} }]{
\includegraphics[width=4.4cm, height=3.6cm]{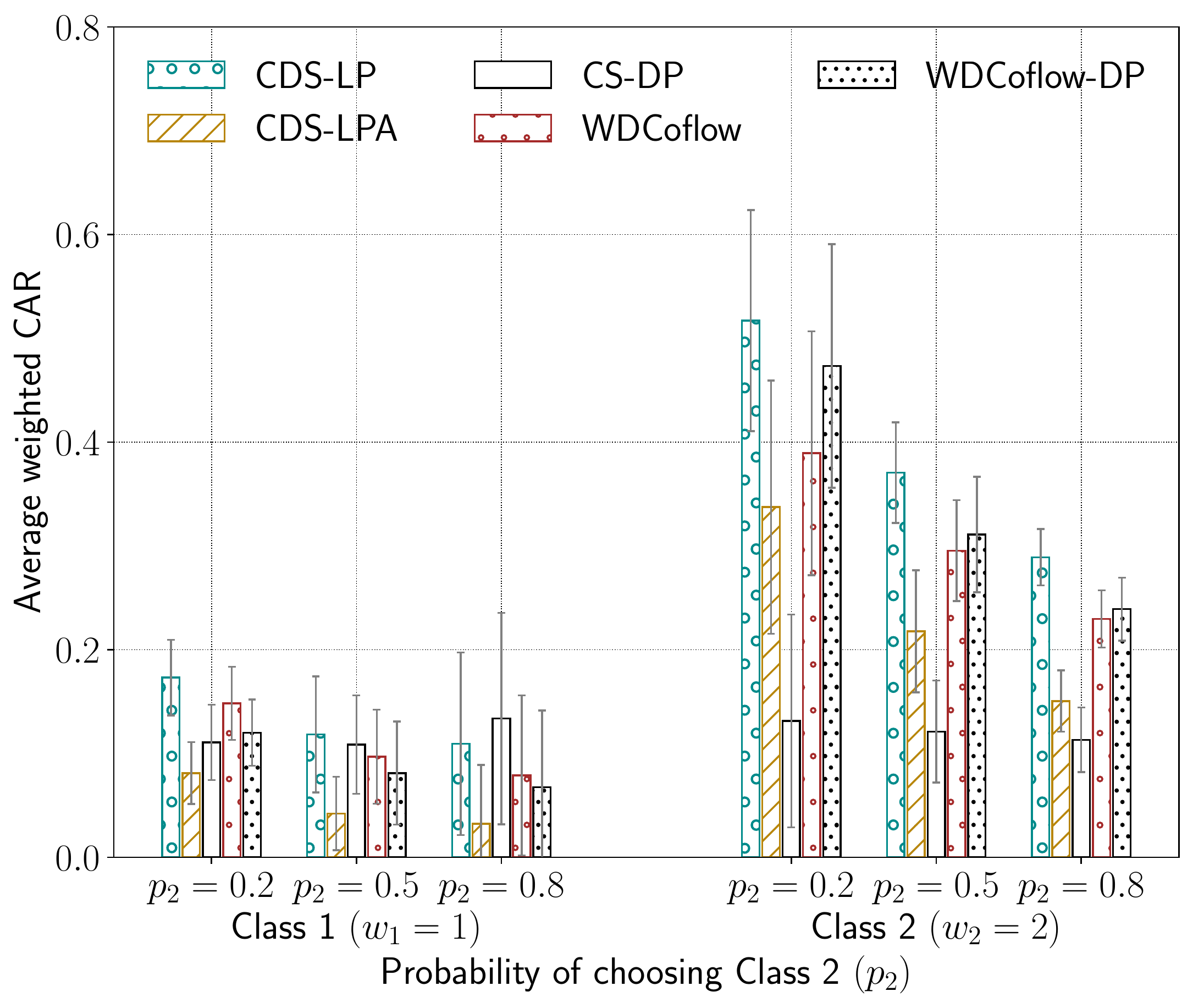}}
\subfloat[{Varying $w_2$ with fixed $p_2$.\label{fig:Off:Syn:ChangeParam:w2} }]{
\includegraphics[width=4.4cm, height=3.6cm]{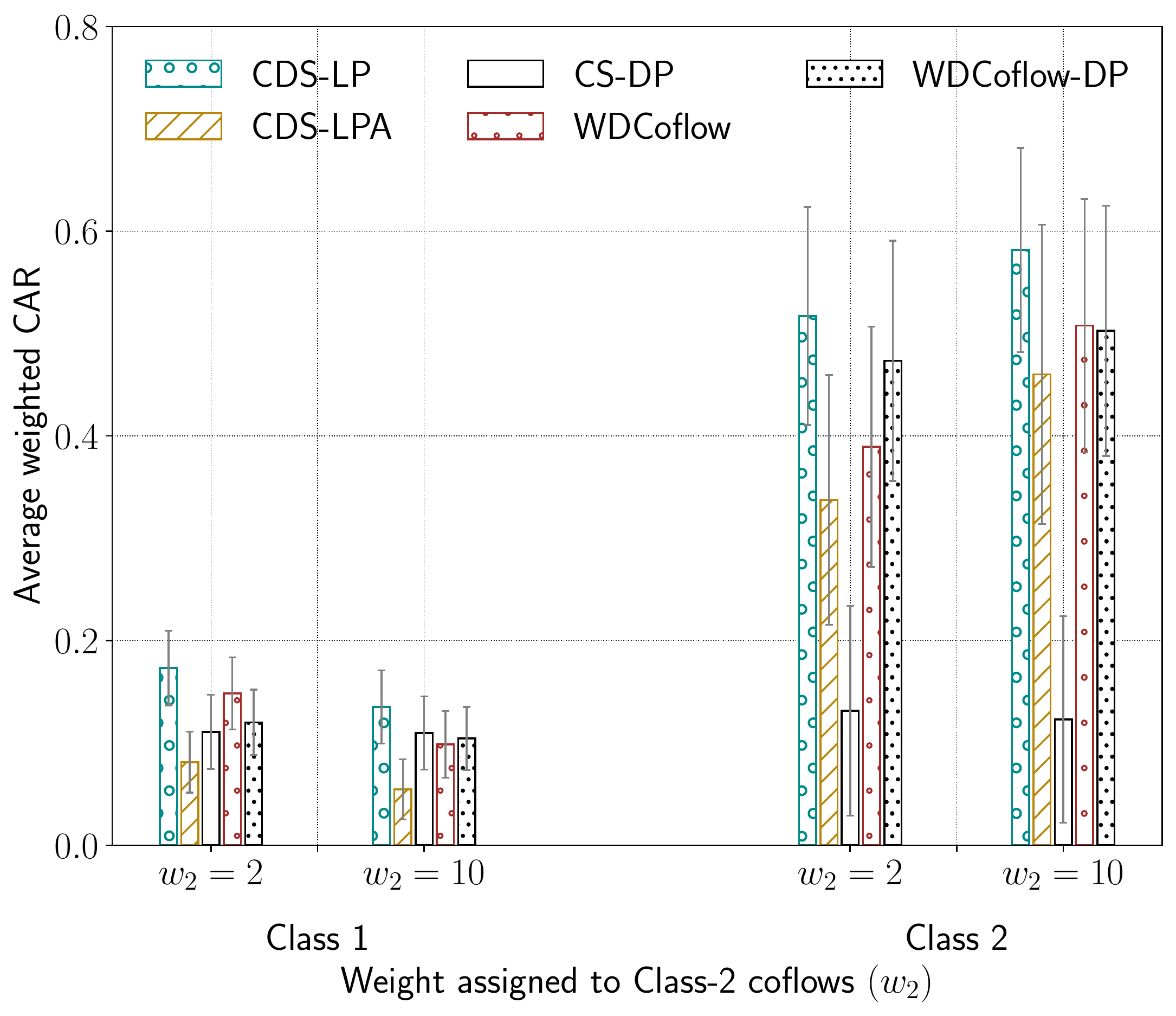}}
\caption{Average per-class CAR with synthetic traffic traces using network configuration $[10,60]$ when (a) varying $p_2$ with fixed $w_2 = 2$ and (b) varying $w_2$ with fixed $p_2 = 0.2$. \label{fig:Off:Syn:ChangeParam}}
\end{figure}

Figs.~\ref{fig:Off:Syn:ChangeParam:p2} and \ref{fig:Off:Syn:ChangeParam:w2} illustrate respectively the per-class WCAR with synthetic traffic traces using network configuration $[10,60]$ when varying $p_2$ (with fixed $w_2$) and $w_2$ (with fixed $p_2$).   We can see that both schedulers  \wdcoflow\ and \wdcoflowDP\  obtain almost the same performance compared to the optimal solution and they handle the priority between classes as \cdslp.  But for  Class $1$, we can see that \csdp\ performs best for  $p_2=0.5$ and $p_2=0.8$.  This means that  \wdcoflow\ and \wdcoflowDP\  take  into account the importance of weight on how to schedule  the coflows.

\paragraph{Real Traffic Traces}\label{sec:Eva:Off:Fb}


Now we use Facebook traces to evaluate  the performance of \wdcoflow\ and \wdcoflowDP. 

\begin{figure}[h]
\centering
\subfloat[{Small-scale networks.\label{fig:Off:Fb:WCAR:Small} }]{
\includegraphics[width=4.4cm, height=3.6cm]{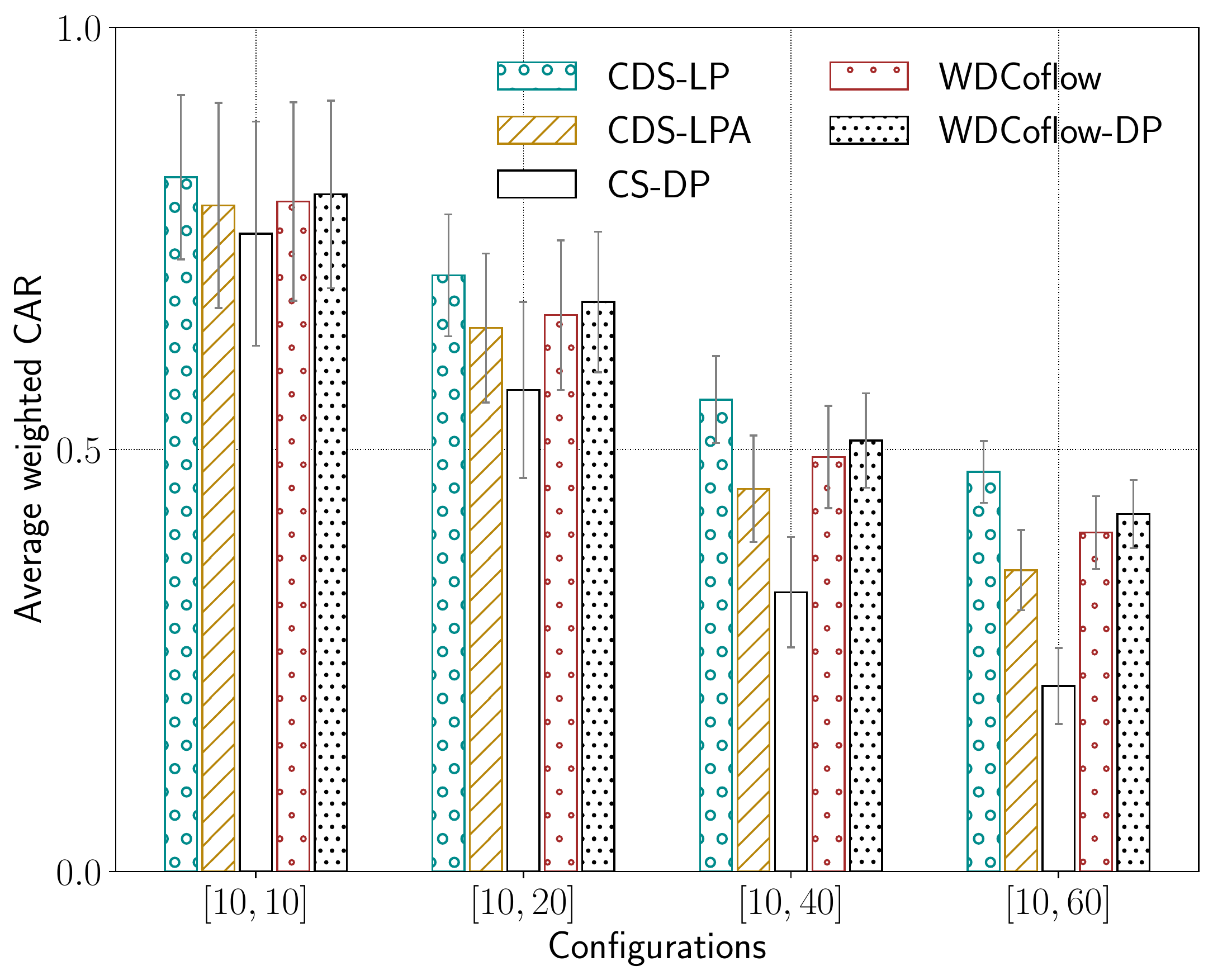}}
\subfloat[{Large-scale networks.\label{fig:Off:Fb:WCAR:Big} }]{
\includegraphics[width=4.4cm, height=3.6cm]{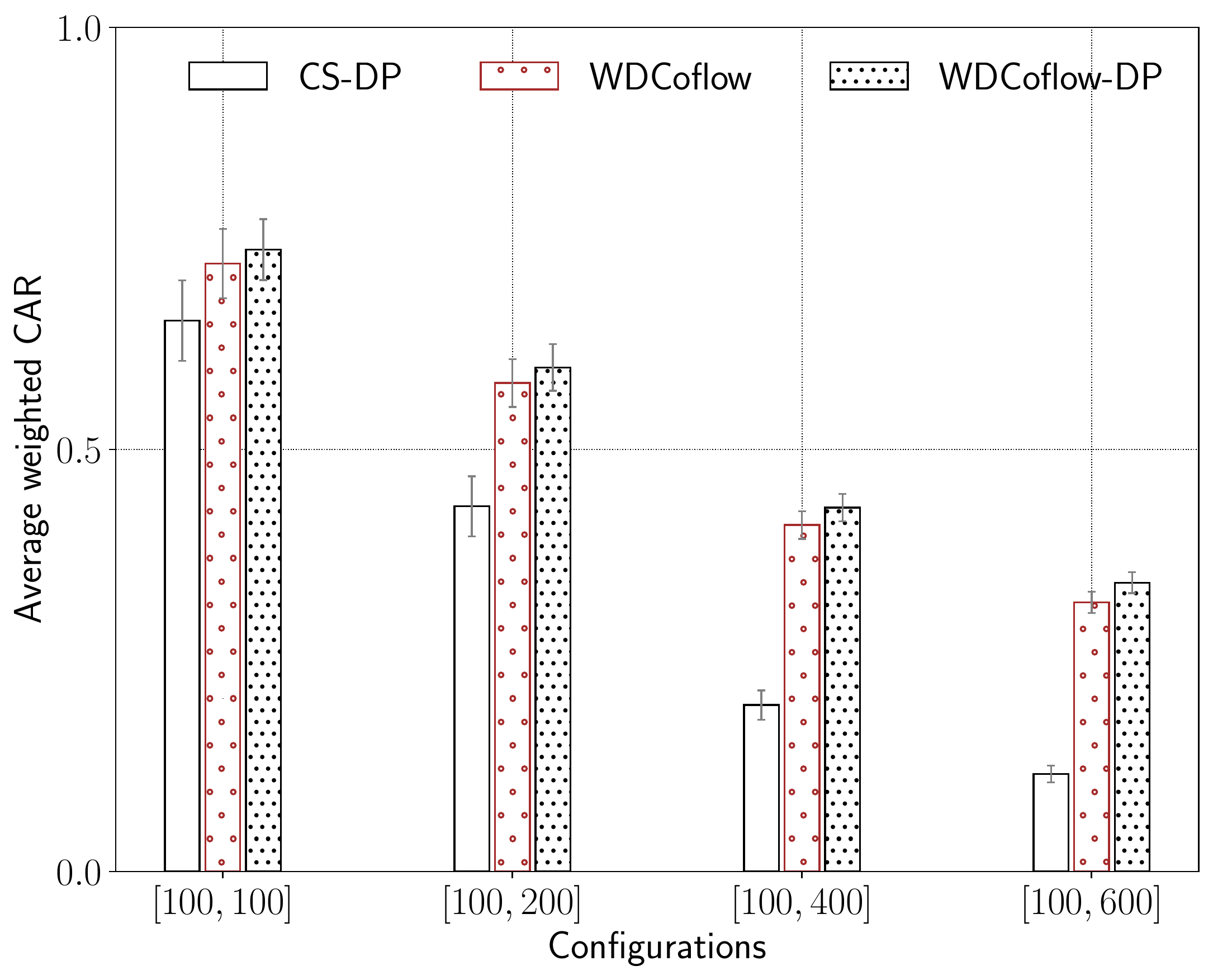}}
\caption{Average WCAR with Facebook traffic traces using (a) small and (b) large-scale networks. The values of $(p_2,w_2)$ are set to $(0.2,2)$. Each point in the x-axis represents the network $[M,N]$. \label{fig:Off:Fb:WCAR}}
\end{figure}

Figs.~\ref{fig:Off:Fb:WCAR:Small} and \ref{fig:Off:Fb:WCAR:Big} show the average WCAR with Facebook traffic traces using small and large-scale networks.  The figures show that  \wdcoflow\ and \wdcoflowDP \ provide near-optimal solutions (the difference is less than $3\%$) while CDS-LPA and CS-DP are far from the optimum of $5\%$ and $8\%$ respectively.   For high load ($N = 60$), \wdcoflow\ and \wdcoflowDP \ lose only $10\%$ and $15\%$ respectively  compared to the optimal but the other two algorithms lose more ground by about $53\%$ compared to the optimal. 
Moreover, the performance gap becomes higher with the increase of the network scale. For instance, with network $[100,100]$, \wdcoflow\ and \wdcoflowDP\ perform around $6.8\%$ and $8.4\%$ better than \csdp\, while with the network $[100,600]$, these gaps become $20\%$ and $22\%$, respectively.

We observe that  \wdcoflow\ and \wdcoflowDP\ approximate the performance of the optimal solution for the small-scale network and perform better than \csdp\  w.r.t. average WCAR. Moreover, the performance gap becomes higher with the increase of the network scale. For instance, with network $[100,100]$, \wdcoflow\ and \wdcoflowDP\ perform around $10\%$ and $13\%$ better than \csdp\, while with network $[100,600]$, these gaps become $331\%$ and $345\%$, respectively.

\begin{figure}[t]
\centering
\subfloat[{Small-scale networks.\label{fig:Off:Fb:ClassCAR:Small} }]{
\includegraphics[width=4.4cm, height=3.6cm]{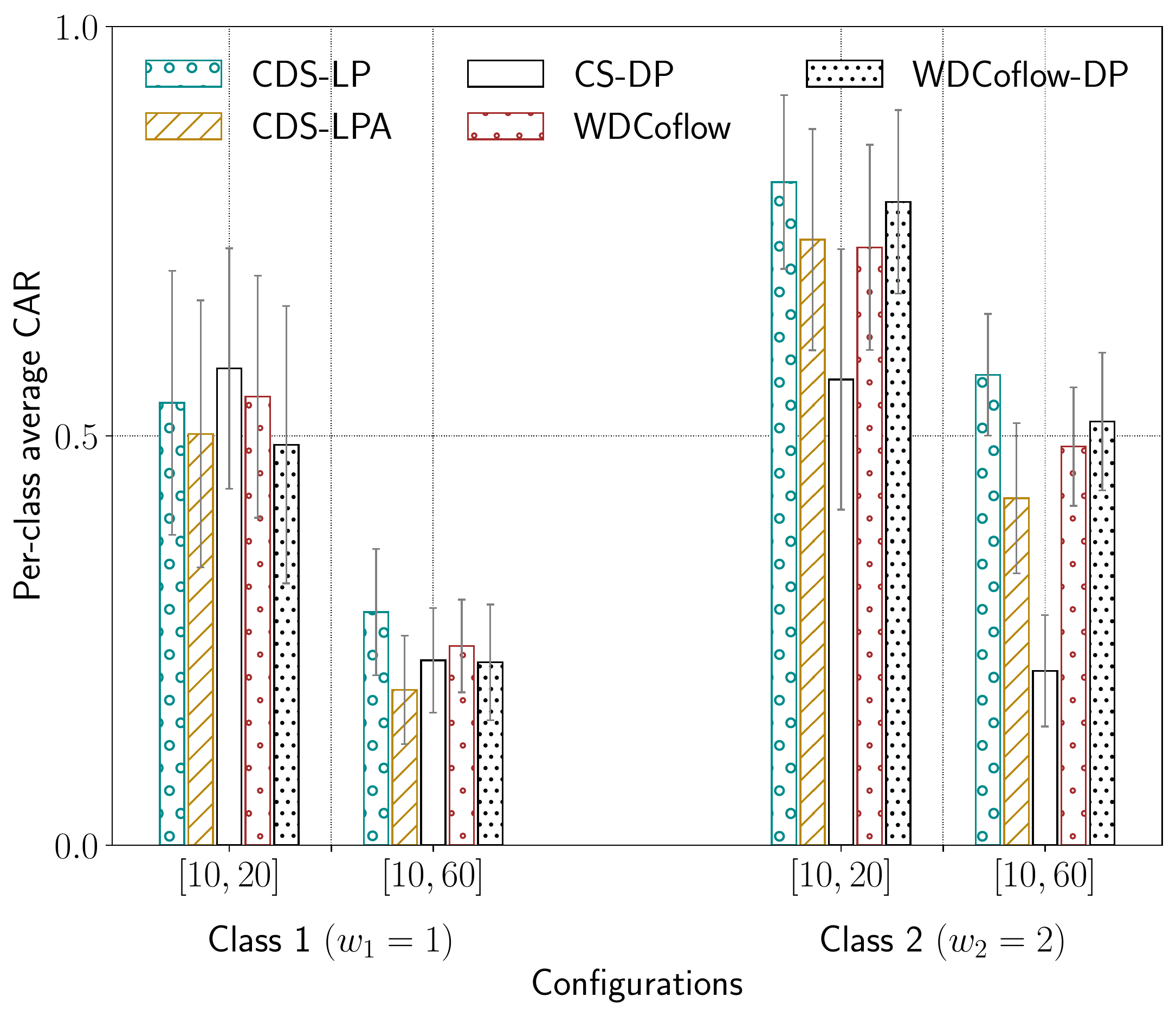}}
\subfloat[{Large-scale networks.\label{fig:Off:Fb:ClassCAR:Big} }]{
\includegraphics[width=4.4cm, height=3.6cm]{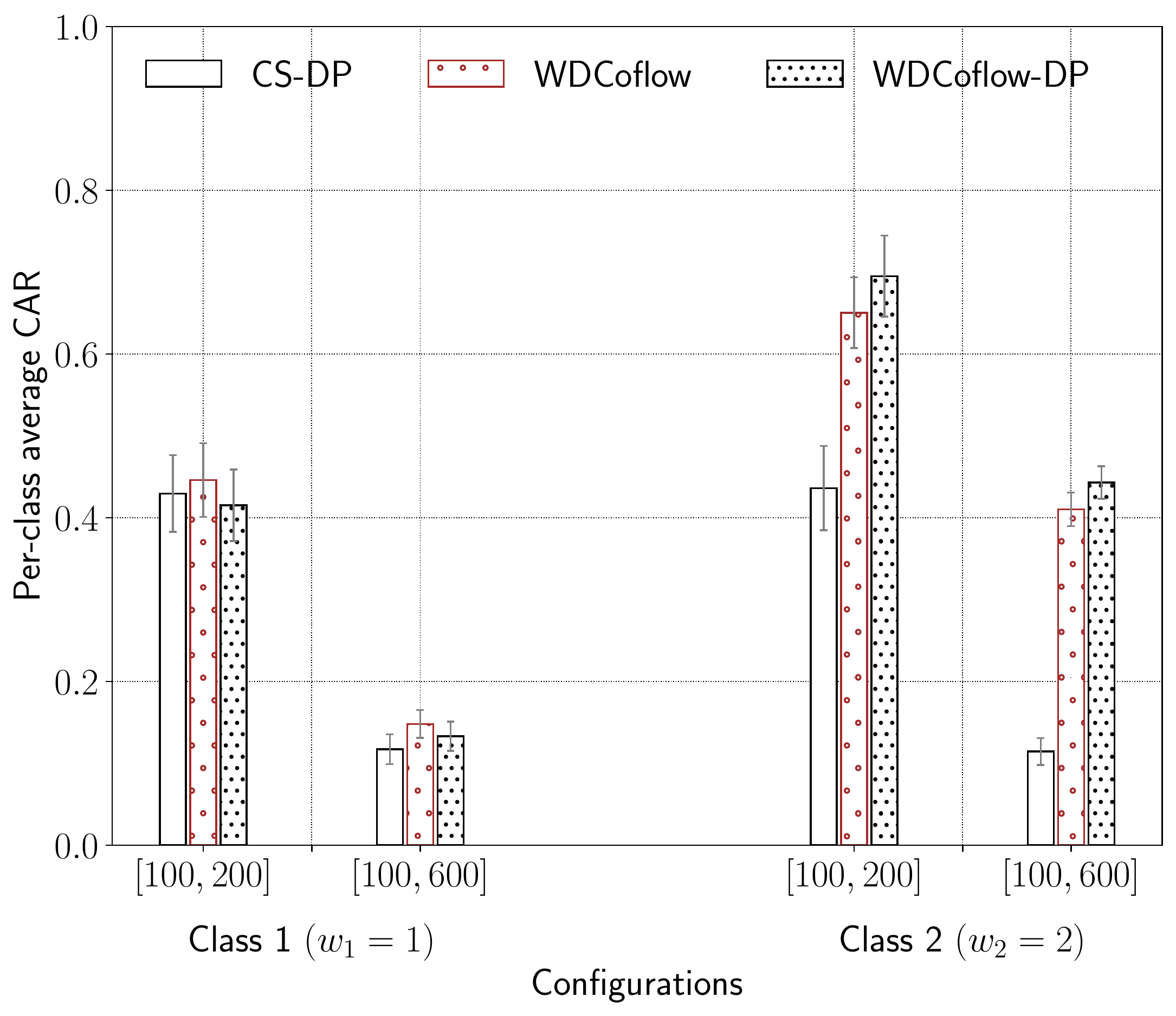}}
\caption{Average per-class CAR with Facebook traffic traces using (a) small-scale and (b) large-scale networks. Each point in the x-axis represents the network $[M,N]$. \label{fig:Off:Fb:ClassCAR}}
\end{figure}

Figs.~\ref{fig:Off:Fb:ClassCAR:Small} and \ref{fig:Off:Fb:ClassCAR:Big} shows the average CAR of each coflow class with Facebook traffic traces using small and large-scale networks, with $(p_2,w_2) = (0.5,2)$.  We can  see that   \wdcoflow\ and \wdcoflowDP\  achieve big improvement for Class $2$, while \csdp\ obtains worse  performance for both classes. The reason
for this is that  \csdp\  schedules coflows only according to
the network conditions, while  \wdcoflow\ and \wdcoflowDP\  consider both network
conditions and coflow weights.  Under   \wdcoflow\ and \wdcoflowDP\,
higher weight coflows have higher priority, thus the average WCAR greatly increases.

\subsubsection{Online Setting\label{sec:Eva:On}}
We now present numerical results comparing
the performance of the online version of \wdcoflow\ and \wdcoflowDP\  against the online version of \csdp. The results are obtained on instances generated using our workload generator, with $50$ machines and $3000$ coflow arrivals. 
Coflow arrivals follow a Poisson process with an average rate of $\lambda$ (coflows/time slot). The arrival rate $\lambda$ varies from $2$ to $10$, and the probability and weight of Class-2 coflows are fixed to respectively $0.5$ and $10$. For the sake of comparison, we have used the greedy allocation algorithm (see the beginning of Sec.~\ref{sec:Evaluation}) to perform the resource allocation after obtaining the $\sigma$-order. For each algorithm the average performance is calculated over $40$ runs with $40$ different instances of the same setting.

\begin{figure}[h]
\centering
\subfloat[{Average WCAR.\label{fig:On:Syn:ChangeLambda:WCAR} }]{
\includegraphics[width=4.4cm, height=3.6cm]{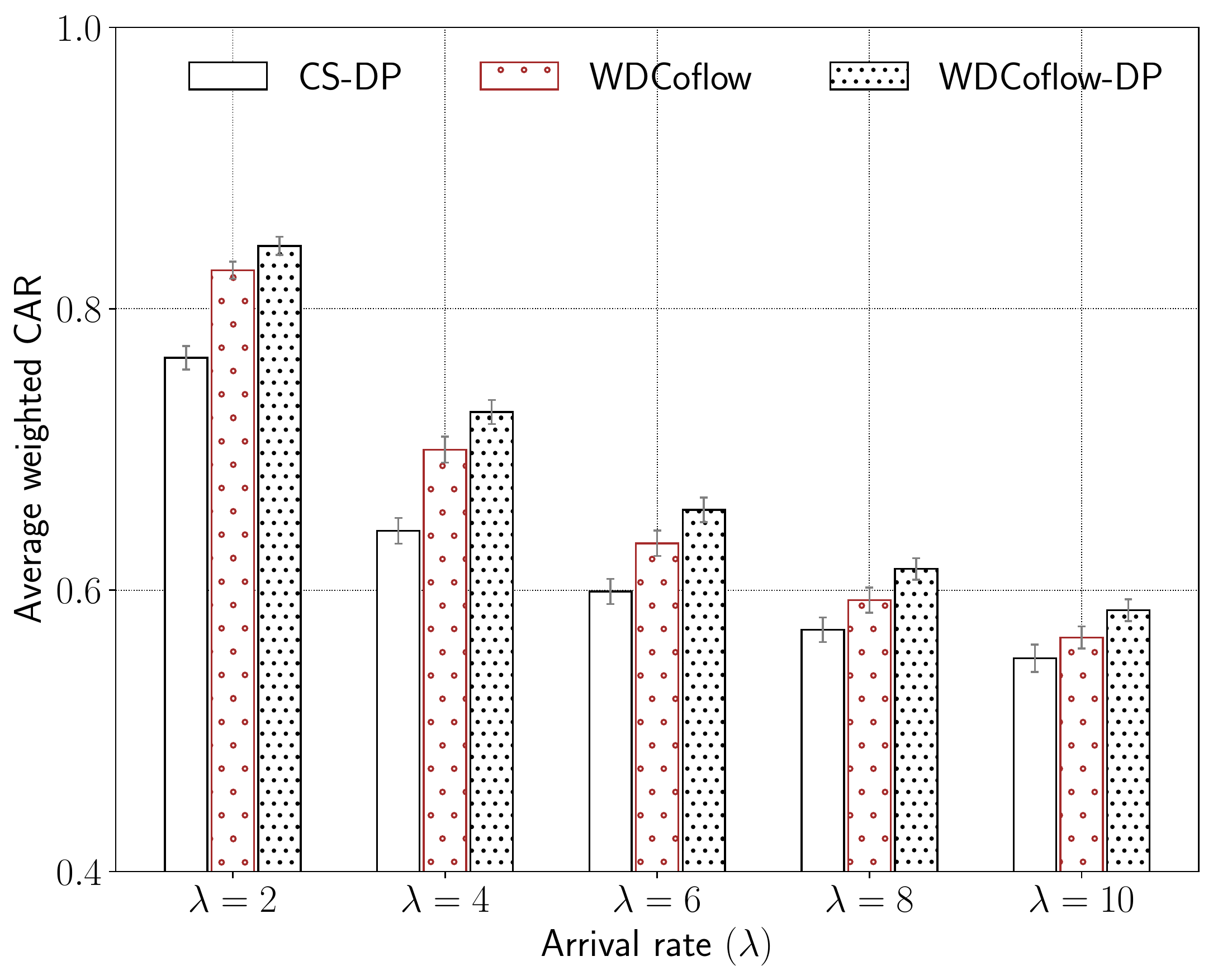}}
\subfloat[{Per-class CAR.\label{fig:On:Syn:ChangeLambda:ClassCAR} }]{
\includegraphics[width=4.4cm, height=3.6cm]{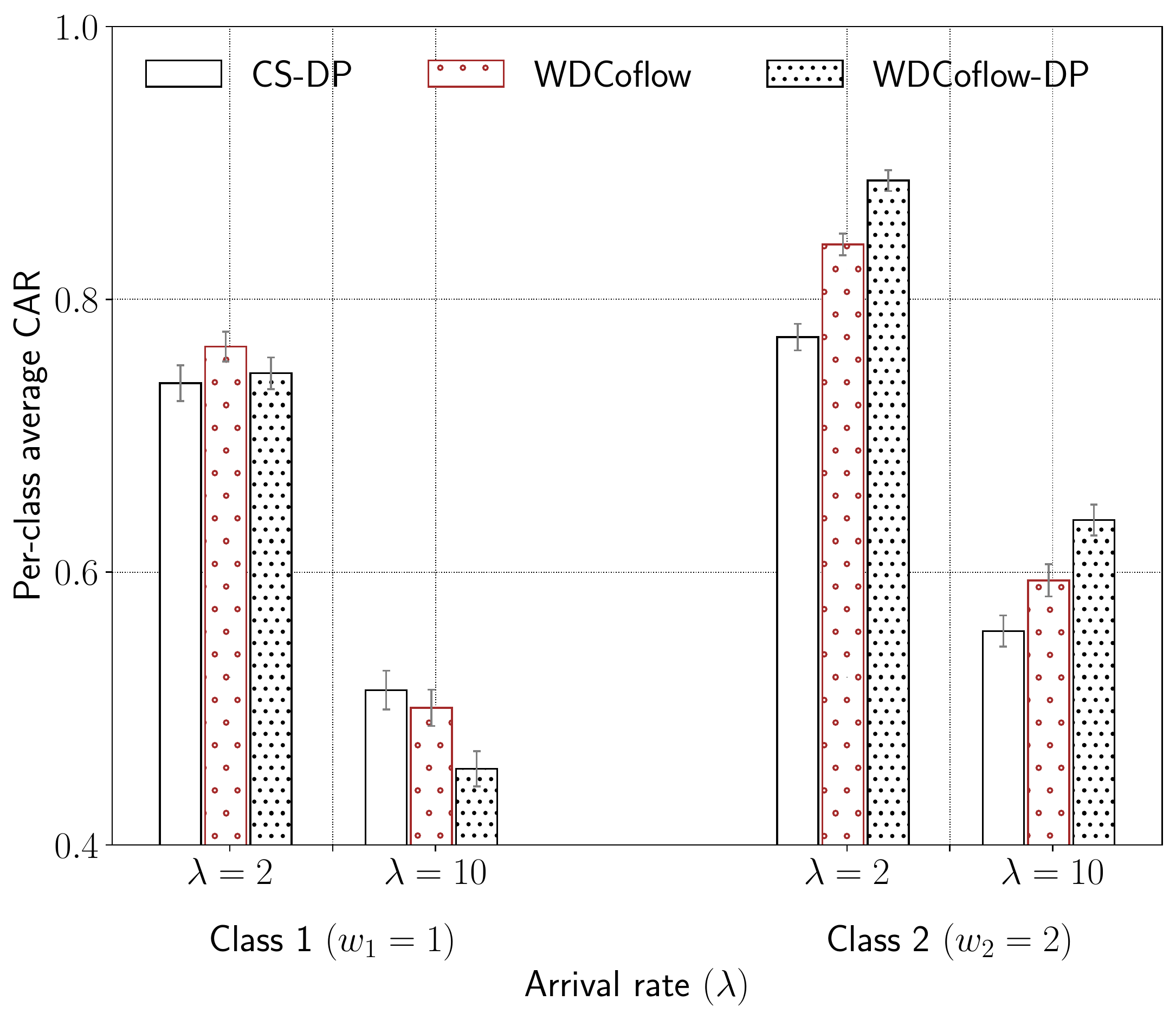}}
\caption{(a) Average WCAR and (b) per-class CAR with synthetic traffic traces when varying $\lambda$ and fixing $p_2=0.5$ and $w_2=2$. \label{fig:On:Syn:ChangeLambda}}
\end{figure}

Fig.~\ref{fig:On:Syn:ChangeLambda:WCAR} and \ref{fig:On:Syn:ChangeLambda:ClassCAR} illustrate the WCAR and per-class CAR of coflows. We observe that  \wdcoflow\ and \wdcoflowDP\ improve  the average WCAR as compared to \csdp. For instance, with $\lambda = 4$, the WCAR improvement of \wdcoflow\ and \wdcoflowDP\ compared to \csdp\ are respectively $7\%$ and $12\%$. 
In addition, they greatly improve the CAR for Class $2$  for all $\lambda$ compared to \csdp\ (see Fig.~\ref{fig:On:Syn:ChangeLambda:ClassCAR}). This shows that our proposed solution consider both network conditions and the importance of coflows to determine the $\sigma$-order. This allows to improve the average CAR and also to differentiate the CAR for a specific target class.

\section{Related Work 
\label{sec:Related-Work}}



\lqt{
In the literature, there has been a stronger emphasis on minimizing the CCT of coflows rather than considering deadline-sensitive scheduling. This discrepancy highlights the relatively lower attention given to deadline scheduling. One of the earlier algorithms that addresses deadline-sensitive coflow scheduling is \varys\ \cite{Chowdhury2014}. \varys\ employs a cascade of coflow admission control and scheduling mechanisms. The scheduler aims to minimize CCT through a combination of strategies, including (\textit{i}) a heuristic for coflow ordering based on the bottleneck's completion time for each coflow and (\textit{ii}) an allocation algorithm that assigns bandwidth to individual flows within each coflow. The rate allocation in \varys\ is designed to approximately align the completion times of all coflows with the bottleneck completion time.


$\mathtt{Chronos}$ \cite{Ma2016} is another heuristic algorithm specifically designed for deadline scheduling. It addresses the issue of flow starvation by allocating residual bandwidth to flows that do not meet their deadlines. The algorithm begins by establishing a priority order among the coflows. Each coflow is then allocated the minimum required bandwidth to meet its individual deadline. If there is insufficient bandwidth available for a particular coflow, it is removed from the allocation and marked for multiplexing. After allocating bandwidth to all flows that meet their deadlines, the remaining bandwidth is distributed proportionally among the remaining coflows based on their demands. This ensures that coflows that cannot fully meet their deadlines still receive a fair share of the available bandwidth.


In \cite{Luo2016}, the authors establish a connection between the problem of deadline scheduling of coflows and the concurrent open shop problem, which is a well-known \textit{NP}-hard problem. They propose a heuristic approach based on the Moore-Hodgson algorithm \cite{Moore1968}, which deals with the case of  single link. A centralized and decentralized version of the heuristic are introduced, namely \csmha\ and $\mathtt{D\textsuperscript{2}\text{-}CAS}$, respectively.

A formal formulation for the deadline scheduling problem including bandwidth allocation of flows is introduced in \cite{Tseng2019}. The CDS maximization problem is cast as an MILP (called \cdslp). In the formulation, time is divided into intervals based on the boundaries set by the coflows deadlines, arranged in increasing order. 
The objective of \cdslp\ is to determine which coflows to accept and the corresponding amount of bandwidth to allocate in each interval to maximize the overall satisfaction of coflow deadlines. The problem takes into account the inherent trade-off between accepting more coflows and allocating sufficient bandwidth to meet their deadlines.
\cdslp\ is proven to be \textit{NP}-hard, indicating that finding an optimal solution is computationally challenging. As an alternative, they propose an approximation algorithm based on LP relaxation, referred to as \cdslpa. \cdslpa\ relaxes the binary variables in the MILP formulation and retains only the coflows that are completely accepted according to the relaxed variables (i.e., their relaxed variables are strictly equal to $1$).

An online heuristic to maximize coflow admissions, while ensuring that their deadlines are met, is presented in \cite{ccgrid20}.  
The authors only focus on comparing their heuristic with \varys, acknowledging that other more efficient algorithms have been developed in the literature, such as those presented in \cite{Tseng2019, Luo2016, Ma2016}.

$\mathtt{MixCoflow}$ \cite{mix22}  addresses the problem of simultaneous optimization of coflows with and without deadline.  
The paper formulates an optimization framework to schedule coflows, with objective to minimize and balance the bandwidth usage of coflows with deadlines, allowing coflows without deadlines to be scheduled as soon as possible.
The framework is fist cast as an ILP, then an equivalent LP problem has been investigated to obtain the optimal solution with lesser computational complexity.  

\cite{Luo23} handles the scenario where the network is overloaded and it becomes impossible to complete all coflows within their respective deadlines. The proposed solution, namely $\mathtt{Poco}$, leverages the observation that certain parallel time-sensitive data applications can tolerate incomplete or partial transmission of their data. $\mathtt{Poco}$ proposes a mechanism to order the coflows  at the limit of the tolerance of each application.

For completeness, we also cite additional works focusing on the problem of minimizing CCT of coflows \cite{ChowdhuryThesis2015,Shaf2018,agarwal2018sincronia,Chen2016,Shi2021,ELITE22} as well as the survey article \cite{Wang2018}. 
Among these, the algorithm \sincronia\ \cite{Agarwal2018} has gained popularity. It addresses CCT minimization by scheduling coflows on network bottlenecks and provides a scheduling order that achieves a $4$-approximation factor.
}

%
%
%
%
%
%
%
%
%



\section{Conclusion and Future Work
\label{sec:Conclusion}}


\lqt{
In this paper, we have presented a novel approach for handling coflow admission control and scheduling in the context of batch processing with deadline constraints. Our algorithm takes advantage of open-shop scheduling techniques to identify a subset of coflows to be scheduled and determines a  $\sigma$ priority order, for efficient execution. By utilizing this $\sigma$-order, coflows are scheduled based on their priority, ensuring effective management of deadlines and improved overall performance.


The experimental evaluation of our algorithms demonstrates promising performance on small-scale networks, where they either match or outperform other existing deadline-sensitive algorithms proposed in prior works. However, the true strength of our approach is revealed on large-scale networks, where it exhibits substantial improvements compared to the existing algorithms. For instance, in an offline setting, our scheme achieves a significantly higher CAR, such as a remarkable $98\%$ increase compared to \csmha. Additionally, our proposed algorithm showcases a remarkable accuracy in prediction: even though the admission control is performed using a CCT approximation with bottleneck ports,  the proposed algorithm ensures that nearly all accepted coflows are able to complete within their assigned deadlines when they are actually scheduled.


This behavior is observed in various scenarios and network settings, including offline and online scenarios, using a wide range of network scales with either synthetic or real traces from the Facebook data set. This  demonstrates the robustness and efficacy of the proposed algorithm when dealing with different situations.


Several extensions of this research line are possible and will be considered for future works. Specifically, the problem of scheduling coflows with incomplete information, e.g., the volume of flows of different coflows. This could occur when the exact volume of a flow of a given coflow is not directly available to the scheduler, but is instead inferred from \textit{a priori} distribution. Understanding how our algorithm performs under such circumstances can provide insights into its robustness and adaptability to uncertain or incomplete information. Finally, issues of starvation and fairness among coflows are also important aspects that have not been addressed in our current work. Future research could focus on developing new algorithms that promote fairness and mitigate the potential for starvation, ensuring equitable treatment of coflows and improving overall system performance.

}



\bibliographystyle{IEEEtran} 
\bibliography{ref_coflow}

\begin{IEEEbiography}[{\includegraphics[width=1in,height=1.25in,clip,keepaspectratio]{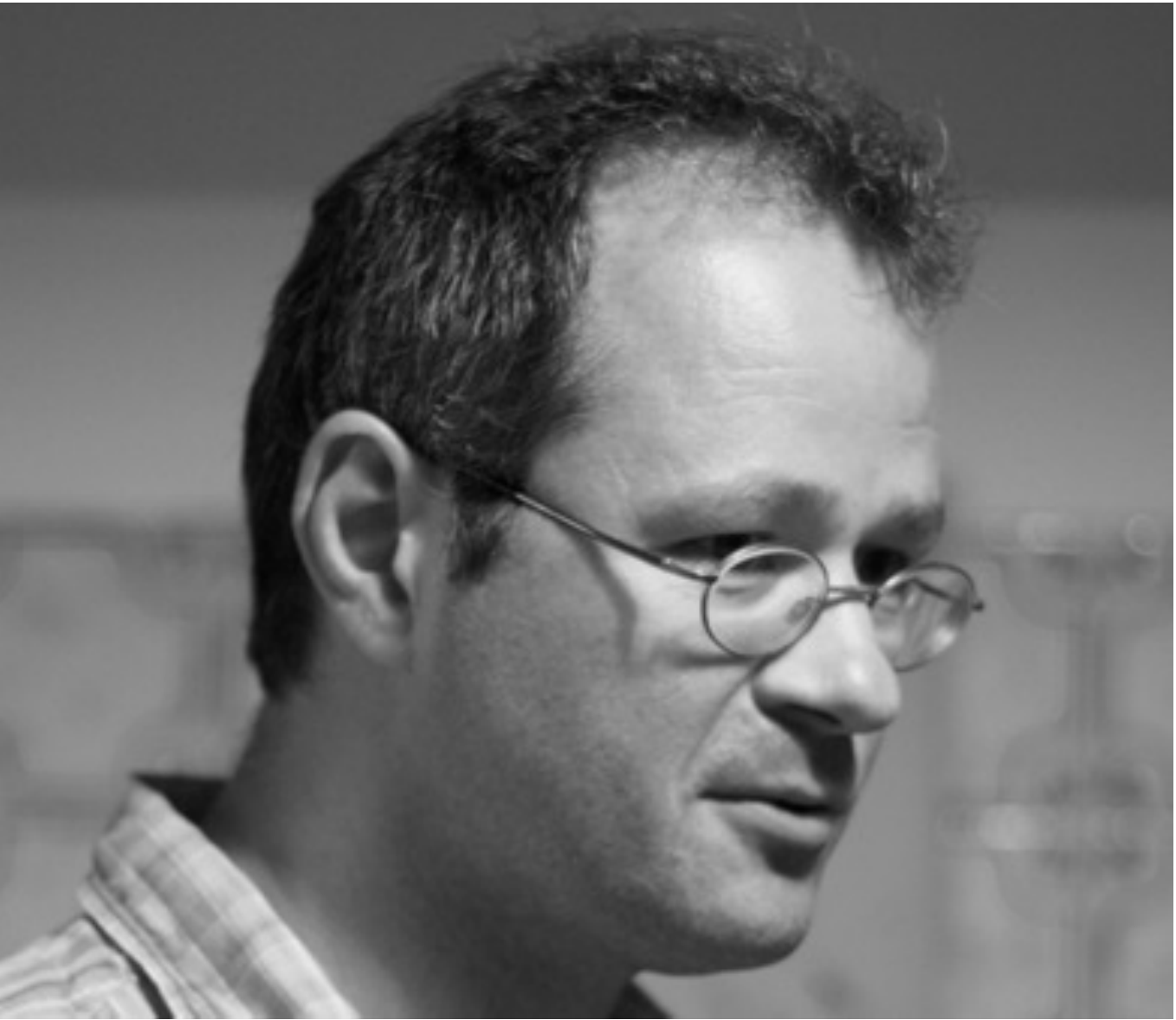}}]{Olivier Brun} is a CNRS research staff member at LAAS, in the SARA group. He graduated from the Institut National des T\'el\' ecommunication (INT, Evry, France) and he was awarded his PhD degree from Universit\' e Toulouse III (France).  His research interests lie in queueing and game theories as well as network optimization.
\end{IEEEbiography}

\vskip -2.8\baselineskip plus -1fil
\begin{IEEEbiography}[{\includegraphics[width=1in, height=1.25in, clip,keepaspectratio]{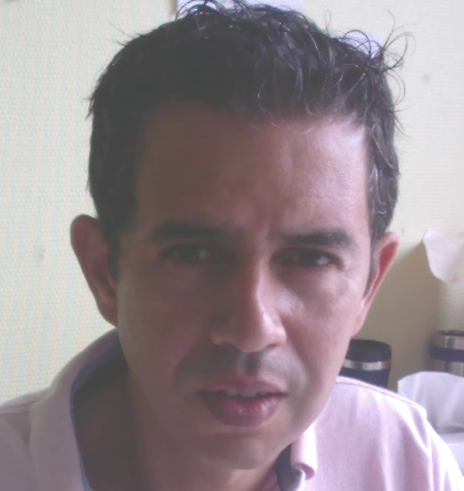}}]{Rachid El-Azouzi}  is a full professor at the University of Avignon. He received his PhD in Applied Mathematics from Mohammed V University in 2000. He joined the National Institute for Research in Computer Science and Control (INRIA), in Sophia Antipolis, where he held positions as a postdoctoral fellow and research engineer. In 2003, he joined the University of Avignon as an associate professor.  His research interests include networked games, resource allocation, wireless networks, complex systems and performance evaluation. 
\end{IEEEbiography}

\vskip -2\baselineskip plus -1fil
\begin{IEEEbiography}[{\includegraphics[width=1in,height=1.25in,keepaspectratio]{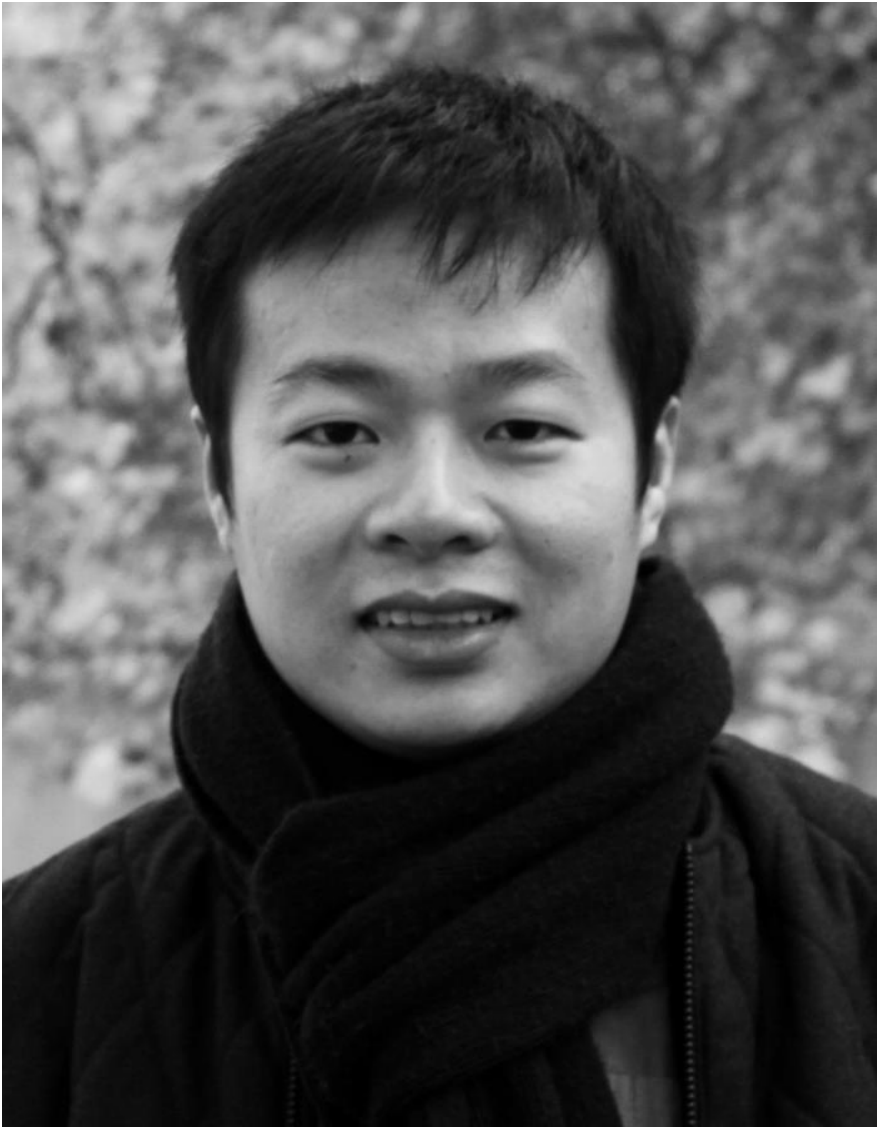}}]{Quang-Trung Luu}is currently a lecturer at Hanoi University of Science and Technology (HUST), Hanoi, Vietnam. He received a Ph.D from CentraleSup\'elec, Paris-Saclay University, France in 2021 (in collaboration with Nokia Bell Labs France). Before joining HUST, he was a postdoctoral fellow at LAAS-CNRS and University of Avignon, France. His research focuses on the optimization of resource management in next-generation communication networks. 
\end{IEEEbiography}

\vskip -2\baselineskip plus -1fil
\begin{IEEEbiography}
    [{\includegraphics[width=1in,height=1.25in,clip,keepaspectratio]{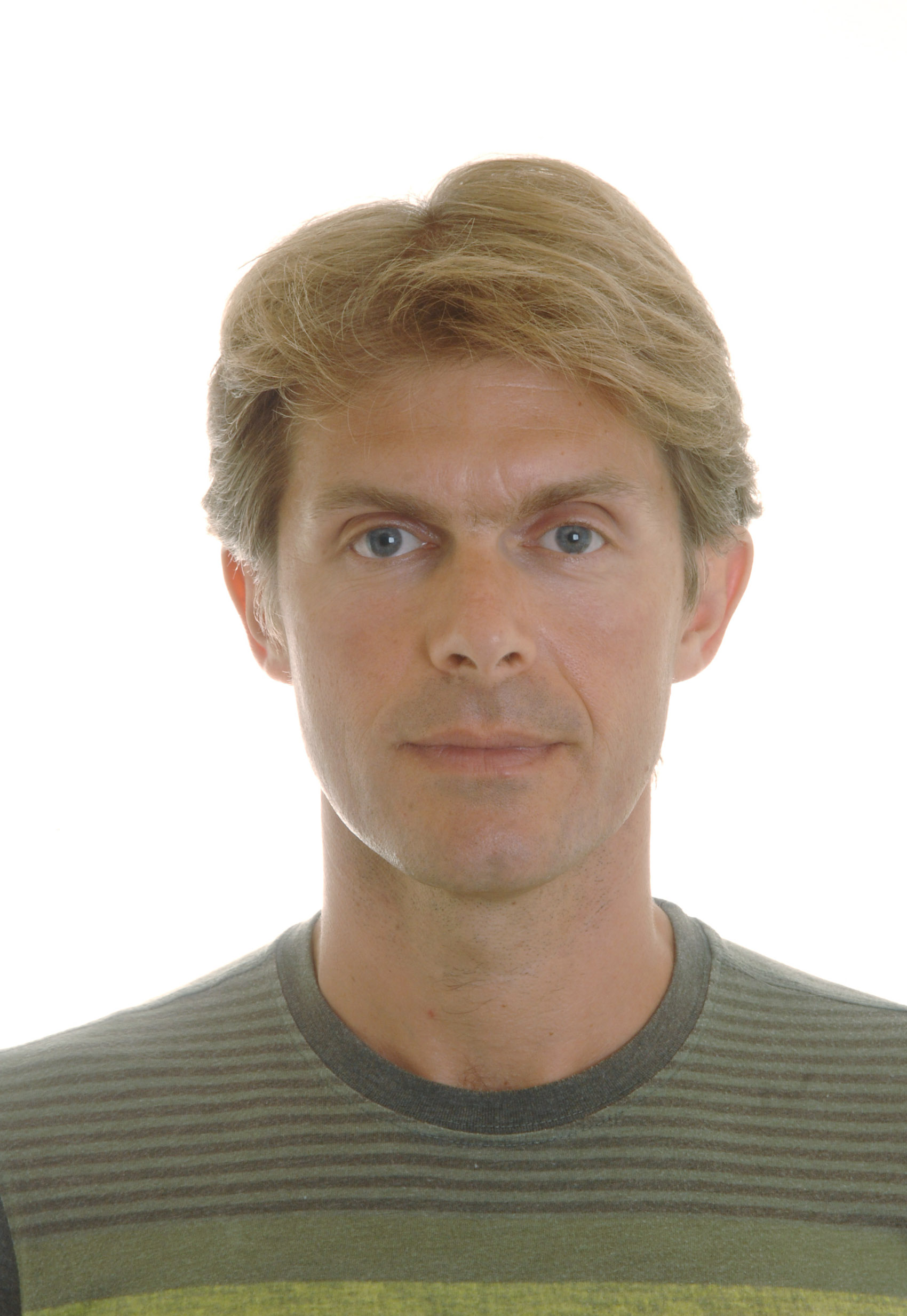}}]
    {Francesco De Pellegrini} received the  MSc 2000, and Ph.D. 2004, University of Padova, Italy, in Information Engineering. He is professor in networking and artificial intelligence at LIA, the Computer Science department of the University of Avignon. Before he was a researcher 
 at Fondazione Bruno Kessler, Italy. He applies algorithms on graphs, stochastic  control, and game theory for the design and the perfomance evaluation of networked systems.
\end{IEEEbiography}

\vskip -2\baselineskip plus -1fil
\begin{IEEEbiography}[{\includegraphics[width=1in,height=1.25in,clip,keepaspectratio]{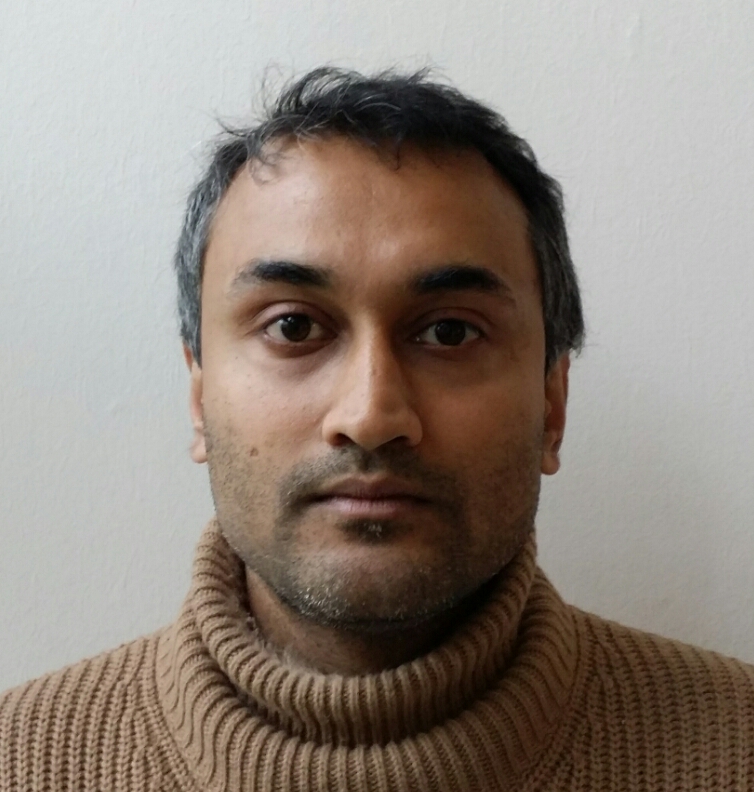}}]{Balakrishna Prabhu} is a CNRS researcher at LAAS-CNRS, Toulouse, France. 
		His research interests are in performance analysis of communication 
		systems using stochastic modelling and game theory. He obtained his PhD 
		from INRIA Sophia Antipolis (France) in 2005 and M.Sc (Engg.) from the 
		IISc (India). Before joining LAAS-CNRS, he did postdoctoral stints at 
		VTT (Finland), CWI, Eurandom and TU/e (The Netherlands).
\end{IEEEbiography}

\vskip -2\baselineskip plus -1fil
\begin{IEEEbiography}[{\includegraphics[width=1in,height=1.75in,clip,keepaspectratio]{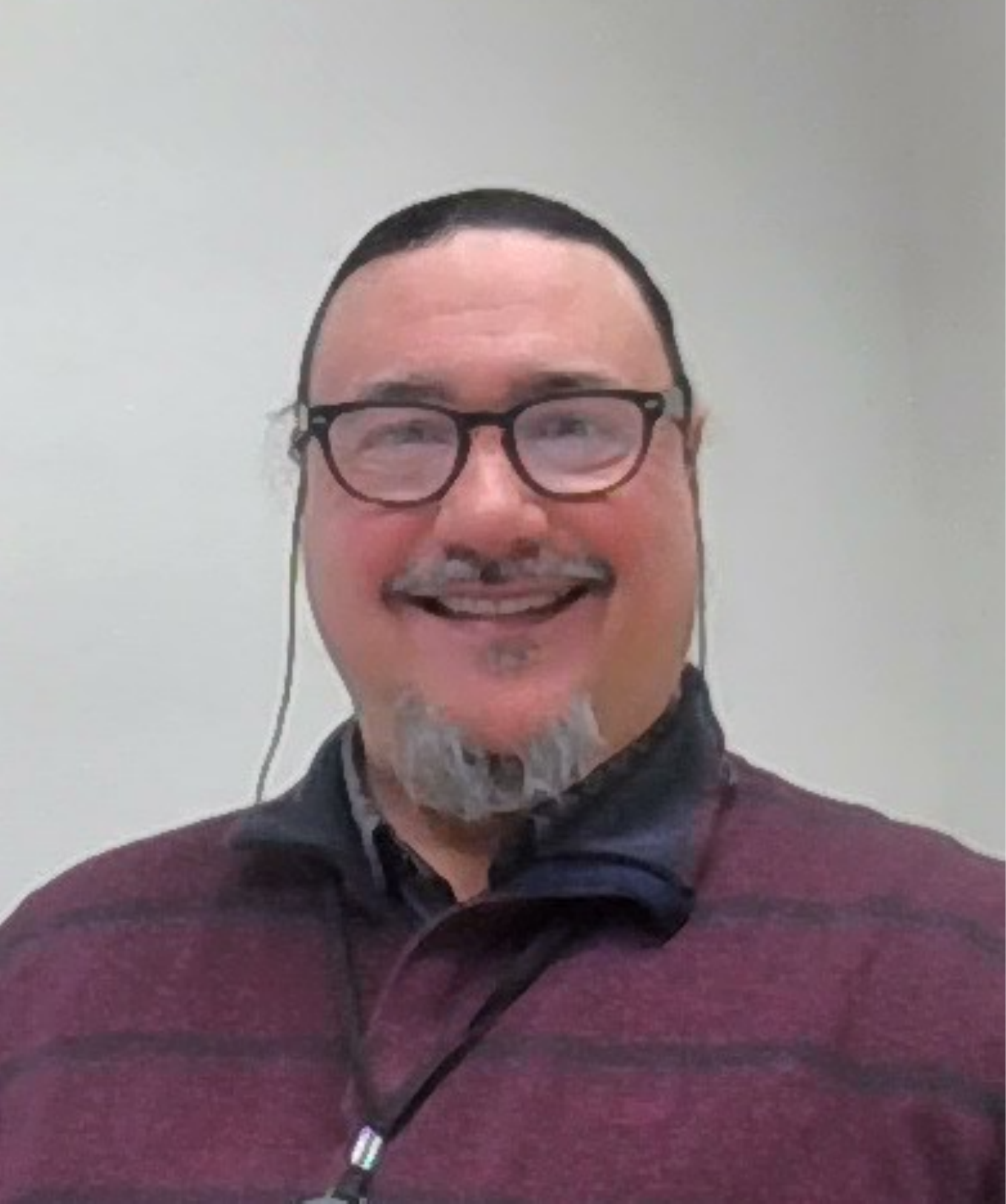}}]{C\'edric Richier } is a research engineer at CNRS and the Avignon University, Avignon, France.   He was awarded his master's degree  in 2012  from the Avignon University. He has worked on several diverse research projects such as social networks, multimedia, data centers and resource allocation.
\end{IEEEbiography}
\end{document}